\documentclass[a4paper,12pt]{article}
\pdfoutput=1
\usepackage{graphicx}
\usepackage{mathtools}
\usepackage{amssymb}
\usepackage{amsfonts}
\usepackage[footnotesize]{caption}
\usepackage[font=scriptsize]{subcaption}
\usepackage{color}
\usepackage{braket}
\usepackage{cite}
\usepackage{hyperref}
\usepackage{url}
\usepackage{multirow}
\usepackage{relsize}
\usepackage{fullpage}
\usepackage{makecell}
\usepackage{blkarray}
\usepackage{fullpage}
\usepackage{bbm,bbold}

\usepackage{url}

\usepackage[utf8]{inputenc}

\newcommand{\newc}{\newcommand}
\newc{\be}{\begin{equation}}
\newc{\ee}{\end{equation}}
\newc{\bea}{\begin{eqnarray}}
\newc{\eea}{\end{eqnarray}}
\newc{\simlt}{~\mbox{\smaller\(\lesssim\)}~}
\newc{\simgt}{~\mbox{\smaller\(\gtrsim\)}~}

\allowdisplaybreaks

\begin{document}

\begin{titlepage}

\begin{flushright} 
IPPP/18/83  \\
\end{flushright} 

\begin{center}
{\bf\Large
\boldmath{
Spontaneous breaking of $SO(3)$ to finite family symmetries with supersymmetry - an $A_4$ model
}
} 
\\[12mm]
Stephen~F.~King$^{\star}$%
\footnote{E-mail: \texttt{king@soton.ac.uk}},
Ye-Ling~Zhou$^{\dagger,\star}$
\footnote{E-mail: \texttt{ye-ling.zhou@durham.ac.uk}},
\\[-2mm]
\end{center}
\vspace*{0.50cm}
\centerline{$^{\star}$ \it
School of Physics and Astronomy, University of Southampton,}
\centerline{\it
Southampton SO17 1BJ, United Kingdom }
\centerline{$^{\dagger}$ \it
Institute for Particle Physics Phenomenology, Department of Physics,}
\centerline{\it
Durham University, South Road, Durham DH1 3LE, United Kingdom}

\vspace*{1.20cm}

\begin{abstract}
{\noindent
We discuss the breaking of $SO(3)$ down to finite family symmetries such as $A_4$, $S_4$ and $A_5$
using supersymmetric potentials for the first time. We analyse in detail the case of 
supersymmetric $A_4$ 
and its finite subgroups $Z_3$ and $Z_2$. We then propose a supersymmetric $A_4$ model
of leptons along these lines, originating from $SO(3)\times U(1)$, which leads to a phenomenologically acceptable pattern of lepton mixing and masses once subleading corrections are taken into account. We also discuss the phenomenological 
consequences of having a gauged $SO(3)$, leading to massive gauge bosons, and show that all domain wall problems are resolved in this model.
}
\end{abstract}
\end{titlepage}

\section{Introduction}

The discovery of neutrino mass and lepton mixing  \cite{nobel} not only represents the first laboratory particle physics beyond the Standard Model (BSM) but also raises additional flavour puzzles such as why the neutrino masses are so small, and why lepton mixing is so large \cite{King:2003jb}. 
Early family symmetry models focussed on continuous non-Abelian gauge theories such as $SU(3)$ \cite{King:2001uz} 
\footnote{$SU(3)$ has recently been considered in extra dimensions \cite{deAnda:2018yfp}.} or $SO(3)$
\cite{King:2005bj}.
Subsequently, non-Abelian discrete symmetries such as $A_4$ 
were introduced, for example to understand the theoretical origin of the observed 
pattern of (approximate) tri-bimaximal lepton mixing \cite{Ma:2001dn,Ishimori:2010au}. 
When supersymmetry (SUSY) is included, the problem of vacuum alignment which is crucial to the success of such 
theories, can be more readily addressed using the flat directions of the potential \cite{Altarelli:2005yp,Altarelli:2005yx,deMedeirosVarzielas:2005qg}.
However, current data involves a non-zero reactor angle and a solar angle which deviate from their tri-bimaximal values \cite{King:2007pr}.
Since, in general, non-Abelian discrete symmetries do not imply either a zero reactor angle or exact tri-bimaximal lepton mixing, these symmetries
are still widely used in current model building \cite{King:2013eh}. 

Although the motivation for non-Abelian discrete symmetries remains strong, there are a few question marks surrounding the use of such symmetries in physics. The first and most obvious question is from where do such symmetries originate? In the Standard Model (SM) we are familiar with the idea of gauge theories being fundamental and robust symmetries of nature, but discrete symmetries seem only relevant to charge conjugation (C), parity (P) and 
time-reversal invariance (T) symmetry \cite{Peccei:1998jv}. 
In supersymmetric (SUSY) models, Abelian discrete symmetries are commonly used to ensure proton stability \cite{Ibanez:1991pr}.
It is possible that the non-Abelian discrete symmetries could arise from some high energy theory such as string theory \cite{Kobayashi:2006wq}, perhaps as a subgroup of the modular group \cite{Altarelli:2005yx,deAdelhartToorop:2011re}
and/or from the orbifolding of extra dimensions \cite{Altarelli:2006kg}.
However, even if such symmetries do arise from string theory, and survive quantum and gravitational corrections \cite{Banks:1991xj}, when they are spontaneously broken they would imply that distinct degenerate vacua exist separated by an energy barrier,
leading to a network of cosmological domain walls which would be in conflict with standard cosmology, and appear to 
``over-close the Universe'' \cite{Zeldovich:1974uw, Kibble:1976sj, Vilenkin:1984ib}.

The problem of domain walls with non-Abelian discrete symmetries such as $A_4$ was discussed in \cite{Riva:2010jm, Antusch:2013toa}
where three possible solutions were discussed: 
\begin{enumerate}
\item to suppose that the $A_4$ discrete symmetry is anomalous,
and hence it is only a symmetry of the classical action and not a full symmetry of the theory, being broken by quantum corrections.
For example this could be due to extending the discrete symmetry to the quark sector such that the symmetry is broken at the quantum level due to the QCD anomaly \cite{Preskill:1991kd}. However, it is not enough to completely solve the problem since this anomaly cannot remove all the vacuum degeneracy \cite{Chigusa:2018hhl};
\item to include explicit $A_4$ breaking terms in the Lagrangian, possibly in the form of Planck scale suppressed higher order operators,
arising from gravitational effects; 
\item to suppose that, in the thermal history of the Universe, the $A_4$ breaking phase transition happens during inflation which effectively dilutes the domain walls, and that the $A_4$ is never restored after reheating following inflation.
\end{enumerate}

An alternative solution to the domain wall problem, which we pursue here, is to suppose that the non-Abelian discrete symmetry arises as a low energy remnant symmetry after the spontaneous breaking of some non-Abelian continuous gauge theory.
This could take place either within the framework of string theory \cite{Banks:2010zn}, or, as in the present paper,
in the framework of quantum field theory (QFT). For example it has been shown how $SO(3)$ can be spontaneously broken to various 
non-Abelian discrete symmetries \cite{Ovrut:1977cn,Etesi:1997jv}. In order to achieve this, a scalar potential was constructed 
such leading to the vaccuum expectation value (VEV) which breaks the continuous gauge symmetry to the discrete symmetry.
The key requirement for having a remnant non-Abelian discrete symmetry seems to be that the scalar field which breaks
the gauge symmetry is in some large irreducible representation (irrep) of the continuous gauge group.

The above approach  \cite{Ovrut:1977cn,Etesi:1997jv} has been applied to flavour models based on non-Abelian continuous gauge symmetries. 
For example, following \cite{Ovrut:1977cn,Etesi:1997jv}, the authors in~\cite{Berger:2009tt} have considered the breaking of gauged 
$SO(3) \to A_4$ by introducing $\underline{7}$-plet of $SO(3)$ with the further breaking of $A_4$ realising tri-bimaximal mixing in a non-SUSY flavour model. 
However, a fine-tuning of around $10^{-2}$ among parameters had to be considered in order to get the correct hierarchy between $\mu$ and $\tau$ masses. 
The problem of how to achieve tri-bimaximal mixing at leading order
from non-Abelian continuous flavour symmetries has also been discussed by other authors~\cite{Koide:2007sr, Wu:2012ria} 
but the problem of determining the required flavon VEVs remains unclear. One idea is to require the electroweak doublets and right-handed fermions to separately transforming under different continuous flavour symmetries, and realise maximal atmospheric mixing from the minimisation of the potential~\cite{Alonso:2013mca, Alonso:2013nca}.
Extended discussions including the breaking of $SU(2)$ and $SU(3)$ to non-Abelian discrete symmetries have been discussed in \cite{Adulpravitchai:2009kd, Grimus:2010ak, Luhn:2011ip, Merle:2011vy, Rachlin:2017rvm,Hernandez:2014lpa} and the phenomenological implications of the breaking of $SU(3)$ flavour symmetry in flavour models has been discussed in \cite{Antusch:2007re, Blankenburg:2012nx}.

The above literature has been concerned with breaking a continuous gauge theory to a 
non-Abelian discrete symmetry {\bf without SUSY}. To date, the problem of how to achieve such a breaking 
{\bf in a SUSY framework} has not been addressed, even though there are many SUSY flavour models in the literature \cite{King:2013eh}.
As stated earlier, the main advantage of such SUSY models is the possibility to achieve vacuum alignment using flat directions of the potential,
which enables some technical simplifications and enhances the theoretical stability of the alignment \cite{Altarelli:2005yp}.
There is also a strong motivation for considering such breaking in a SUSY framework, in order to make 
contact with {\bf SUSY flavour models} \cite{King:2013eh}.
In addition, the usual motivations for embedding the non-Abelian discrete symmetry into a gauge theory also apply
in the SUSY context as well, namely:
\begin{itemize}
\item To provide a natural explanation of the origin of non-Abelian discrete flavour symmetries in SUSY flavour models. 
\item To avoid the domain wall problem of SUSY flavour models,
since the non-Abelian discrete flavour symmetry 
is just an approximate effective residual symmetry arising from the breaking of the continuous symmetry.
When the approximate discrete symmetry is broken it
does not lead to domain walls.
\end{itemize}
Finally, if the continuous symmetry is gauged, there is the phenomenological motivation that:
\begin{itemize}
\item The breaking of {\bf gauged flavour symmetries} to finite non-Abelian flavour symmetries implies new massive 
gauge bosons in the spectrum, with possibly observable phenomenological signatures. For instance, SUSY $SO(3)\to A_4$ will lead to three degenerate gauge bosons plus their superpartners.
\end{itemize}

In the present paper, motivated by the above considerations, 
we discuss the breaking of a continuous {\bf SUSY} gauge theory to a 
non-Abelian discrete symmetry using a potential which 
{\bf preserves SUSY}. As stated above, this is the first time that such a symmetry breaking has been 
discussed in the literature, and the formalism developed here may be applied to the numerous SUSY flavour models in the literature
\cite{King:2013eh}.
For example, 
we discuss the breaking of $SO(3)$ down to finite family symmetries such as $A_4$, $S_4$ and $A_5$
using supersymmetric potentials for the first time. In particular,
we focus in detail on the breaking of SUSY $SO(3)$ to $A_4$,
with SUSY preserved by the symmetry breaking. We further show how the $A_4$ may be subsequently broken to smaller residual 
symmetries $Z_3$ and $Z_2$, still preserving SUSY, which may be used to govern the mixing patterns in the charged lepton and neutrino sectors,
leading to a predictive framework. We then present an explicit SUSY $SO(3)\times U(1)$ model of leptons which uses this symmetry breaking pattern and show that it leads to a phenomenologically acceptable pattern of lepton mixing and masses. 
Finally we discuss the phenomenological 
consequences of having a gauged $SO(3)$, leading to massive gauge bosons, and show that all domain wall problems are resolved in such models.

The layout of the remainder of the paper is then as follows.
In section~\ref{sec:SO3} we discuss the spontaneous breaking of $SO(3)$ to finite non-Abelian symmetries 
such as $A_4$, $S_4$ and $A_5$ with supersymmetry.
In section~\ref{sec:A4} we discuss the further breaking of $A_4$ to residual $Z_3$ and $Z_2$ symmetries,
showing how it may be achieved from a supersymmetric $SO(3)$ potential.
In section~\ref{sec:model} we construct in detail a supersymmetric $A_4$ model along these lines,
originating from $SO(3) \times U(1)$, 
and show that it leads to a phenomenologically acceptable pattern of lepton mixing and masses,
once subleading corrections are taken into account. 
Within this model, we also discuss the phenomenological 
consequences of having a gauged $SO(3)$, leading to massive gauge bosons, and show that all domain wall problems are resolved. Section~\ref{conclusion} concludes the paper.
The paper has three appendices.
In Appendix~\ref{sec:CG} we list the Clebsch-Gordan coefficients of $SO(3)$ which are used in the paper.
In Appendix~\ref{app:minimisation} we display explicitly the solutions of the superpotential minimisation.
In Appendix~\ref{app:Z2_correction} we show the deviation from the $Z_2$-invariant vacuum.

\section{Spontaneous breaking of $SO(3)$ to finite non-Abelian symmetries 
$A_4$, $S_4$ and $A_5$ with supersymmetry \label{sec:SO3}}

The key point to break $SO(3)$ to non-Abelian discrete symmetries is introducing a high irrep of $SO(3)$ and require it gain a non-trivial VEV. In this section, after a brief review of $SO(3)$, we discuss how to break $SO(3)$ to $A_4$ by introducing a $\underline{7}$-plet, and then generalise our discussion to $SO(3) \to S_4$ and $A_5$. 

\subsection{The $SO(3)$ group}

The rotation group $SO(3)$ is one of the most widely used Lie groups in physics and mathematics. It is generated by three generators $\tau^1$, $\tau^2$ and $\tau^3$. Each element can be expressed by 
\begin{eqnarray}
g_{\{\alpha^a\}} = \exp \Big( \sum_{a=1,2,3}\alpha^a \tau^a \Big) = \mathbb{1} + \sum_{a=1,2,3}\alpha^a \tau^a + \frac{1}{2}\Big( \sum_{a=1,2,3}\alpha^a \tau^a \Big)^2 + \cdots \,.
\label{eq:SO3_transformation}
\end{eqnarray}
In the fundamental three dimensional (3d) space, the generators are represented as
\begin{eqnarray}
\tau^1 = \left(
\begin{array}{ccc}
 0 & -1 & 0 \\
 1 & 0 & 0 \\
 0 & 0 & 0 \\
\end{array}
\right) \,, \quad
\tau^2 = \left(
\begin{array}{ccc}
 0 & 0 & -1 \\
 0 & 0 & 0 \\
 1 & 0 & 0 \\
\end{array}
\right) \,, \quad
\tau^3 = \left(
\begin{array}{ccc}
 0 & 0 & 0 \\
 0 & 0 & -1 \\
 0 & 1 & 0 \\
\end{array}
\right) \,.
\label{eq:generator}
\end{eqnarray}

Each irrep of $SO(3)$ has $2p+1$ dimensions and we denote it as a $\underline{2p\!+\!1}$-plet. Each $\underline{2p\!+\!1}$-plet can be represented as a rank-$p$ tensor $T_{i_1 i_2 ... i_p}$ in the 3d space. This tensor is symmetric and traceless, 
\begin{eqnarray}
\phi_{...i_a...i_b...} = \phi_{...i_b...i_a...}\,, \ \ 
\sum_{i_a=i_b=1}^3 \phi_{...i_a...i_b...} = 0 \,, 
\label{eq:symmetric_traceless}
\end{eqnarray} 
for any $a, b \leqslant p$. 
It transforms under $SO(3)$ as 
\begin{eqnarray}
\phi_{i_1 i_2 ... i_p} \to O_{i_1 j_1} O_{i_2 j_2} \cdots O_{i_p j_p} \phi_{j_1 j_2 ... j_p} \,,
\end{eqnarray}
where $O$ is transformation matrix corresponding to the element $g_{\{\alpha^\alpha\}}$ in the 3d space, and it is always a $3 \times 3$ real orthogonal matrix. Here and in the following, doubly repeated indices are summed. 

Products of two irreps can be reduced as $\underline{2p\!+\!1} \times \underline{2q\!+\!1} = \underline{2|p\!-\!q|\!+\!1} + \underline{2|p\!-\!q|\!+\!3} + \cdots + \underline{2(p\!+\!q)\!+\!1}$ and the Clebsch-Gordan coefficients are given in Appendix \ref{sec:CG}.

\subsection{$SO(3) \to$ non-Abelian discrete symmetries}

$SO(3)$ can be spontaneously broken to other non-Abelian discrete symmetries by introducing different high irreps. Ref.~\cite{Etesi:1997jv} gives an incomplete list of subgroups which could be obtained after the relevant irrep get a VEV. For instance, some of those subgroup obtained by irreps up to $\underline{13}$ are shown in Table~\ref{tab:subgroup}. The minimal irrep for $SO(3) \to S_4$ is a $\underline{9}$-plet, while that for $SO(3) \to A_5$ is a $\underline{13}$-plet. Applying a $\underline{9}$-plet flavon $\rho$ and a $\underline{13}$-plet flavon $\psi$, respectively, we will realise these breakings in a SUSY framework in the following.

\begin{table}[h!]
\newcommand{\tabincell}[2]{\begin{tabular}{@{}#1@{}}#2\end{tabular}}
  \centering
  \begin{tabular}{llllllll}\hline\hline
  irrep & $\underline{1}$ & $\underline{3}$ & $\underline{5}$ & $\underline{7}$ & $\underline{9}$ & $\underline{11}$ & $\underline{13}$ \\\hline
  subgroups & 
  $SO(3)$ & 
  \multirow{2}{15mm}{$SO(2)$ \\ $SO(3)$} & 
  \multirow{3}{15mm}{$Z_2 \times Z_2$ \\ $SO(2)$ \\ $SO(3)$} &
  \multirow{6}{15mm}{$\mathbf{1}$ \\ $A_4$ \\ $Z_3$ \\ $D_4$ \\ $SO(2)$ \\ $SO(3)$} &
  \multirow{1}{15mm}{$S_4$} &
  \multirow{1}{15mm}{$\,$} &
  \multirow{3}{15mm}{$\mathbf{1}$ \\ $A_4$ \\ $S_4$ \\ $A_5$}  
 \\[21mm]
 \\\hline\hline
\end{tabular}
  \caption{\label{tab:subgroup} The not systematical stabiliser subgroups in the low-dimensional irreducible representations of the group $SO(3)$ \cite{Etesi:1997jv}. }
\end{table}

\subsubsection{$SO(3) \to A_4$}
The simplest irrep to break $SO(3) \to A_4$ is using a $\underline{7}$-plet \cite{Ovrut:1977cn,Etesi:1997jv}. In this work, we introduce a $\underline{7}$-plet flavon $\xi$ to achieve this goal. 
In the 3d flavour space, it is represented as a rank-3 tensor $\xi_{ijk}$, which satisfies the requirements in Eq.~\eqref{eq:symmetric_traceless}, i.e.,
\begin{eqnarray}
&&\xi_{ijk}=\xi_{jki}=\xi_{kij}=\xi_{ikj}=\xi_{jik}=\xi_{kji} \,, \ \ 
\xi_{iik} =0\,. 
\label{eq:rank3}
\end{eqnarray}
Constrained by Eq.~\eqref{eq:rank3}, there are 7 free components of $\xi$, which can be chosen as 
\begin{eqnarray}
\xi_{111},\,\xi_{112},\,\xi_{113},\,\xi_{123},\,\xi_{133},\,\xi_{233},\,\xi_{333}\,.
\end{eqnarray}

\begin{figure}[h!]
\centering
\includegraphics[scale=.8]{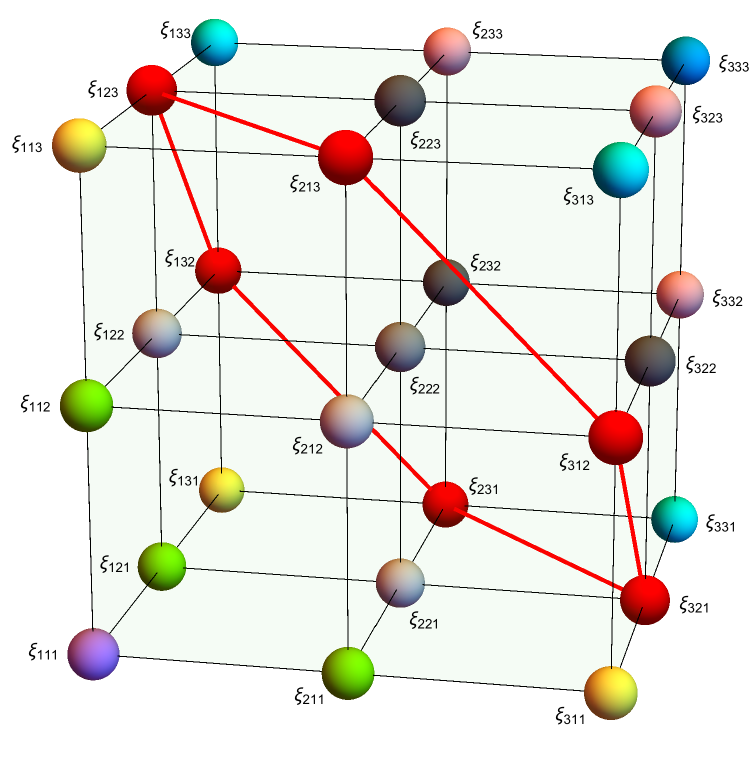}
\caption{\label{fig:xi_VEV} A geometrical description of the $\underline{7}$-plet $\xi_{ijk}$ as a tank-3 tensor with $i,j,k=1,2,3$. Points in the same colour represent the identical components, e.g., $\xi_{112} = \xi_{121} = \xi_{211}$ all in green, etc. As a traceless tensor, points in grey are dependent upon the rest, e.g., $\xi_{122} = \xi_{212} = \xi_{221} = -\xi_{111}-\xi_{133}$. These properties leave only 7 independent components, showing in 7 different colours. For the $A_4$-invariant VEV, only those in red, $\xi_{123}=\xi_{132}=\xi_{231} = \xi_{213} = \xi_{312} = \xi_{321}$, take non-zero values. }
\end{figure}

For the $A_4$ symmetry, we work in the Ma-Rajasekaran (MR) basis, where the generators $s$ and $t$ in the 3d irreducible representation are given by 
\begin{eqnarray}
g_s=\left(
\begin{array}{ccc}
 1 & 0 & 0 \\
 0 & -1 & 0 \\
 0 & 0 & -1 \\
\end{array}
\right)\,,\quad
g_t=\left(
\begin{array}{ccc}
 0 & 0 & 1 \\
 1 & 0 & 0 \\
 0 & 1 & 0 \\
\end{array}
\right)\,.
\end{eqnarray}
The $A_4$-invariant VEV, satisfying 
\begin{eqnarray}
(g_s)_{i i'}(g_s)_{j j'}(g_s)_{k k'} \langle \xi_{i' j' k'} \rangle = \langle \xi_{i j k} \rangle \,,\nonumber\\
(g_t)_{i i'}(g_t)_{j j'}(g_t)_{k k'} \langle \xi_{i' j' k'} \rangle = \langle \xi_{i j k} \rangle \,,
\end{eqnarray} 
is given by 
\begin{eqnarray}
\langle \xi_{123} \rangle \equiv \frac{v_\xi}{\sqrt{6}} \,, \ \ \langle \xi_{111} \rangle = \langle \xi_{112} \rangle = \langle \xi_{113} \rangle = \langle \xi_{133} \rangle = \langle \xi_{233} \rangle = \langle \xi_{333} \rangle = 0 \,.
\label{eq:xi_VEV}
\end{eqnarray}
The VEV of $\xi$ is geometrically shown in Fig.~\ref{fig:xi_VEV}. 

The discussion of $SO(3) \to A_4$ has been given in Refs.~\cite{Ovrut:1977cn,Etesi:1997jv,Berger:2009tt}. The main idea is constructing flavon potential and clarifying the $A_4$-invariant one in Eq.~\eqref{eq:xi_VEV} to be the minimum of the potential, where $v_\xi$ is determined by the minimisation. This idea cannot be directly applied to supersymmetric flavour models. In the later case, the flavon potential is directly related to the flavon superpotential
\begin{eqnarray}
V_f = \sum_i \left| \frac{\partial w_f}{\partial \phi_i} \right|^2 + \cdots \,, 
\label{eq:flavon_potential}
\end{eqnarray}
where $\phi_i$ represent any scalars in the theory, and the dots are negligible soft breaking terms and D-terms for the fields charged under the gauge group. This potential is more constrained than the non-supersymmetric version. If the minimisation of the superpotential $\partial w_f / \partial \phi_i = 0$ has a solution, the minimisation of the potential $\partial V_f / \partial \phi_i = 0$ is identical to the minimisation of the superpotential. Since most flavour models have been built in SUSY, it is necessary to consider if $SO(3) \to A_4$ can be achieved in SUSY. 

In order to break $SO(3)$ to $A_4$, we introduce two driving fields $\xi^d_{\underline{1}} \sim \underline{1}$, $\xi^d_{\underline{5}} \sim \underline{5}$ and consider the following superpotential terms
\begin{eqnarray}
w_\xi = \xi^d_{\underline{1}} \left(c_1 (\xi \xi)_{\underline{1}} - \mu^2_\xi \right) + c_2 \big( \xi^d_{\underline{5}} (\xi \xi)_{\underline{5}} \big)_{\underline{1}} \,,
\end{eqnarray}
where $c_1$ and $c_2$ are complex dimensionless coefficients. As required \cite{Altarelli:2005yx}, the driving fields do not gain non-zero VEVs, realised by imposing $U(1)_R$ charges. 
Minimisation of the potential is identical to the minimisation of the flavon superpotential respecting to the driving fields as follows, 
\begin{eqnarray}
\frac{\partial w_\xi}{\partial \xi^d_{\underline{1}}} &=&  c_1 (\xi \xi)_{\underline{1}} - \mu^2_\xi = 0 \,, \label{eq:w_xi_singlet} \\
\frac{\partial w_\xi}{\partial \xi^d_{\underline{5}}} &=& c_2 (\xi \xi)_{\underline{5}} = 0 \,. \label{eq:w_xi_quintet} 
\end{eqnarray}
The explicit expressions of Eqs.~\eqref{eq:w_xi_singlet} and \eqref{eq:w_xi_quintet} are listed in Appendix~\ref{app:minimisation}. 
Taking the $A_4$-invariant VEV to Eqs.~\eqref{eq:w_xi_singlet} and \eqref{eq:w_xi_quintet}, we see that Eq.~\eqref{eq:w_xi_quintet} is automatically satisfied and Eq.~\eqref{eq:w_xi_singlet} leads to $\langle \xi_{123} \rangle = \pm \mu_\xi/\sqrt{6 c_1}$. Therefore, the $A_4$ symmetry is consistent with the vacuum solution obtained from the minimisation of the superpotential. 

We need to check {\bf the uniqueness of $A_4$} since it is not clear if $A_4$ is the only symmetry after $SO(3)$ breaking. 
We assume there is another vacuum solution $\langle \xi \rangle'$, which has an infinitesimal deviation from the $A_4$-invariant one, $\langle \xi \rangle' = \langle \xi \rangle + \delta \xi$. Eqs.~\eqref{eq:w_xi_singlet} and \eqref{eq:w_xi_quintet} must also be satisfied for $\langle \xi \rangle'$. Directly taking them into account, we get the constraints on $\delta \xi$. Straightforwardly, we obtain 
\begin{eqnarray}
\delta \xi_{123} = \delta \xi_{111} = \delta \xi_{333} = 0 \,, \quad  \delta \xi_{112} + \delta \xi_{233} = 0\,,\end{eqnarray}
leaving only three unconstrained parameters $\delta \xi_{112}$, $\delta \xi_{113}$ and $\delta \xi_{133}$. The unconstrained perturbation parameters $\delta \xi$ can be rotated away if we consider a $SO(3)$ basis transformation, $g_{\{\alpha^a \}}$ in Eq.~\eqref{eq:SO3_transformation}
with $\alpha^1 = {\sqrt{\frac{3c_1}{2}} \delta \xi_{113}}/{\mu }$, $\alpha^2 = {\sqrt{\frac{3c_1}{2}} \delta \xi_{112}}/{\mu }$, $\alpha^3 = -{\sqrt{\frac{3c_1}{2}} \delta \xi_{133}}/{\mu }$ and the generators $\tau^i$ being given in Eq.~\eqref{eq:generator}. 
Therefore, $\langle \xi \rangle'$ also preserves the $A_4$ symmetry and the shift from $\langle \xi \rangle$ to $\langle \xi \rangle'$ corresponds to only a basis transformation of $SO(3)$. Such a basis transformation has no physical meaning. We conclude that the minimisation equation of the superpotential, i.e., Eqs.~\eqref{eq:w_xi_singlet} and \eqref{eq:w_xi_quintet}, uniquely breaks $SO(3)$ to $A_4$. 

\subsubsection{$SO(3)\to S_4$}

For the $S_4$ symmetry, the generators in the 3d irreducible space are given by $g_s$, $g_t$ and 
\begin{eqnarray}
g_u=-\left(
\begin{array}{ccc}
 0 & 1 & 0 \\
 1 & 0 & 0 \\
 0 & 0 & 1 \\
\end{array}
\right)\,.
\end{eqnarray}

In the 3d flavour space, the $\underline{9}$-plet $\rho$ is represented as a rank-4 tensor $\rho_{ijkl}$. 
Constrained by Eq.~\eqref{eq:symmetric_traceless}, there are 9 free components of $\rho$, which can be chosen as 
\begin{eqnarray}
\rho_{1111},\,\rho_{1112},\,\rho_{1113},\,\rho_{1123},\,\rho_{1133},\,\rho_{1233},\,\rho_{1333},\,\rho_{2333},\,\rho_{3333}\,.
\end{eqnarray}
In order to require the VEV $\langle \rho \rangle$ invariant under the $S_4$ symmetry. The following constraints are required,
\begin{eqnarray}
(g_s)_{i i'}(g_s)_{j j'}(g_s)_{k k'} (g_s)_{l l'} \langle \rho_{i' j' k' l'} \rangle = \langle \rho_{i j k l} \rangle \,,\nonumber\\
(g_t)_{i i'}(g_t)_{j j'}(g_t)_{k k'} (g_t)_{l l'} \langle \rho_{i' j' k' l'} \rangle = \langle \rho_{i j k l} \rangle \,,\nonumber\\
(g_u)_{i i'}(g_u)_{j j'}(g_u)_{k k'} (g_u)_{l l'} \langle \rho_{i' j' k' l'} \rangle = \langle \rho_{i j k l} \rangle \,,
\end{eqnarray} 
which are equivalent to 
\begin{eqnarray}
&&\langle \rho_{1111} \rangle = \langle \rho_{3333} \rangle = -2 \langle \rho_{1133} \rangle\,, \nonumber\\
&&\langle \rho_{1112} \rangle = \langle \rho_{1113} \rangle = \langle \rho_{1123} \rangle = \langle \rho_{1233} \rangle = \langle \rho_{1333} \rangle = \langle \rho_{2333} \rangle = 0 \,.
\label{eq:S4_VEV}
\end{eqnarray}

Follwing a similar procedure but replacing the $\underline{7}$-plet $\xi$ by a $\underline{9}$-plet $\rho$, we succeed to break $SO(3)$ to $S_4$ in SUSY by introducing two driving fields $\rho^d_{\underline{1}} \sim \underline{1}$ and $\rho^d_{\underline{5}} \sim \underline{5}$. The flavon superpotential is constructed as
\begin{eqnarray}
w_\rho = \rho^d_{\underline{1}} \left( \mu^2_\rho - c_{\rho1} (\rho \rho)_{\underline{1}} \right) + c_{\rho2} \rho^d_{\underline{5}} (\rho \rho)_{\underline{5}} \,.
\end{eqnarray}
Minimisation respect to the driving fields gives rise to 
\begin{eqnarray}
\frac{\partial w_\rho}{\partial \rho^d_{\underline{1}}} &=& \mu_\rho^2 - c_{\rho1}(\rho \rho)_{\underline{1}} = 0 \,, \label{eq:w_singlet_rho} \\
\frac{\partial w_\rho}{\partial \rho^d_{\underline{5}}} &=& c_{\rho2} (\rho \rho)_{\underline{5}} = 0 \,. \label{eq:w_quintet_rho} 
\end{eqnarray}
Taking Eq.~\eqref{eq:S4_VEV} to the above equations, we see that Eq.~\eqref{eq:w_quintet_rho} is automatically satisfied and Eq.~\eqref{eq:w_singlet_rho} leads to $\langle \rho_{1133} \rangle = \pm \mu_\rho/\sqrt{30c_{\rho1}}$. 

{\bf The uniqueness of $SO(3) \to S_4$.} 
We vary $\rho$ away from the $S_4$-invariant VEV, $\rho \to \langle \rho \rangle + \delta \rho$ and require that Eqs.~\eqref{eq:w_singlet_rho} and \eqref{eq:w_quintet_rho} are still satisfied. Then, we will get the constraints on $\delta \rho$, which are straightforwardly expressed as 
\begin{eqnarray}
\delta \rho_{1111} = \delta \rho_{1123} = \delta \rho_{1133} = \delta \rho_{1233} = \delta \rho_{3333} = 0 \,, \quad  \delta \rho_{1113} + \delta \rho_{1333} = 0\,,\end{eqnarray}
leaving only three unconstrained parameters $\delta \rho_{1112}$, $\delta \rho_{1113}$ and $\delta \rho_{2333}$. The unconstrained perturbation parameters $\delta \rho$ can be rotated away if we consider a $SO(3)$ basis transformation, 
$g_{\{\alpha^a \}} = \mathbb{1}_{3 \times 3} + \alpha^a \tau^a$
with $\alpha^1 = {\sqrt{\frac{6c_{\rho1}}{5}} \delta \rho_{1112}}/{\mu }$, $\alpha^2 = {\sqrt{\frac{6c_{\rho1}}{5}} \delta \rho_{1113}}/{\mu }$, $\alpha^3 = -{\sqrt{\frac{6c_{\rho1}}{5}} \delta \rho_{2333}}/{\mu }$. Therefore, Eqs.~\eqref{eq:w_singlet_rho} and \eqref{eq:w_quintet_rho} uniquely break $SO(3)$ to $S_4$. 

\subsubsection{$SO(3)\to A_5$}

For the $A_5$ symmetry, the generators in the 3d irreducible space are given by $g_s$, $g_t$ and 
\begin{eqnarray}
g_w=-\frac{1}{2} \left(
\begin{array}{ccc}
 -1 & b_{2} & b_{1} \\
 b_{2} & b_{1} & -1 \\
 b_{1} & -1 & b_{2} \\
\end{array}
\right)\,,
\end{eqnarray}
where $b_{1}=\frac{1}{2} (\sqrt{5}-1)$ and $b_{2}=\frac{1}{2} (-\sqrt{5}-1)$. 

The $\underline{13}$-plet $\psi$ in the 3d flavour space is represented as a rank-6 tensor $\psi_{ijklmn}$. 
Constrained by Eq.~\eqref{eq:symmetric_traceless}, there are 13 free components of $\psi$, which can be chosen to be 
\begin{eqnarray}
&\psi_{111111},\,\psi_{111112},\,\psi_{111113},\,\psi_{111123},\,\psi_{111133},\,\psi_{111233},\,\psi_{111333},\,\psi_{112333},\, \\ &\psi_{113333},\,\psi_{123333},\,\psi_{133333},\,\psi_{233333},\,\psi_{333333}\,.
\end{eqnarray}
In order to require the VEV $\langle \psi \rangle$ invariant under the $S_4$ symmetry. The following constraints are required,
\begin{eqnarray}
(g_s)_{i i'}(g_s)_{j j'}(g_s)_{k k'} (g_s)_{l l'} (g_s)_{m m'} (g_s)_{n n'} \langle \psi_{i' j' k' l' m' n'} \rangle = \langle \psi_{i j k l m n} \rangle \,,\nonumber\\
(g_t)_{i i'}(g_t)_{j j'}(g_t)_{k k'} (g_t)_{l l'} (g_t)_{m m'} (g_t)_{n n'}  \langle \psi_{i' j' k' l' m' n'} \rangle = \langle \psi_{i j k l m n} \rangle \,,\nonumber\\
(g_w)_{i i'}(g_w)_{j j'}(g_w)_{k k'} (g_w)_{l l'} (g_w)_{m m'} (g_w)_{n n'}  \langle \psi_{i' j' k' l' m' n'} \rangle = \langle \psi_{i j k l m n} \rangle \,.
\end{eqnarray} 
They are equivalent to 
\begin{eqnarray}
&\langle \psi_{111111} \rangle = \langle \psi_{333333} \rangle,\ 
\langle \psi_{111133} \rangle = \frac{7 \sqrt{5}-5}{10} \langle \psi_{111111} \rangle, \ 
\langle \psi_{113333} \rangle = \frac{-7 \sqrt{5}-5}{10} \langle \psi_{111111} \rangle\,, \nonumber\\
&\langle \psi_{111112} \rangle = \langle \psi_{111113} \rangle = \langle \psi_{111123} \rangle = \langle \psi_{111233} \rangle = 0\,, \nonumber\\
&\langle \psi_{111333} \rangle = \langle \psi_{112333} \rangle = \langle \psi_{133333} \rangle = \langle \psi_{233333} \rangle = 0 \,.
\label{eq:A5_VEV}
\end{eqnarray}

In order to break $SO(3)$ to $A_5$, we introducing two driving fields $\psi^d_{\underline{1}} \sim \underline{1}$ and $\psi^d_{\underline{9}} \sim \underline{9}$, instead of $\underline{5}$. The flavon superpotential is constructed as
\begin{eqnarray}
w_\psi = \psi^d_{\underline{1}} \left( \mu^2_\psi - c_{\rho1} (\psi \psi)_{\underline{1}} \right) + c_{\psi2} \psi^d_{\underline{9}} (\psi \psi)_{\underline{9}} \,.
\end{eqnarray}
Minimisation respect to the driving fields gives rise to 
\begin{eqnarray}
\frac{\partial w_\psi}{\partial \psi^d_{\underline{1}}} &=& \mu_\psi^2 - c_{\psi1}(\psi \psi)_{\underline{1}} = 0 \,, \label{eq:w_singlet_psi} \\
\frac{\partial w_\psi}{\partial \psi^d_{\underline{9}}} &=& c_{\psi2} (\psi \psi)_{\underline{9}} = 0 \,. \label{eq:w_nonet_psi} 
\end{eqnarray}
Taking Eq.~\eqref{eq:A5_VEV} to the above equations, we see that Eq.~\eqref{eq:w_nonet_psi} is automatically satisfied and Eq.~\eqref{eq:w_singlet_psi} leads to $\langle \psi_{111111} \rangle = \pm \mu_\psi/(4\sqrt{21c_{\psi1}})$. 

{\bf The uniqueness of $SO(3) \to A_5$.} 
We vary $\psi$ away from the $A_5$-invariant VEV, $\psi \to \langle \psi \rangle + \delta \psi$ and require that Eqs.~\eqref{eq:w_singlet_psi} and \eqref{eq:w_nonet_psi} are still satisfied. Then, we will get the constraints on $\delta \psi$, 
\begin{eqnarray}
\delta \psi_{111111} = \delta \psi_{111133} = \delta \psi_{113333} = \delta \psi_{333333} = 0 \,, \nonumber\\ 
\delta \psi_{111112} =  \sqrt{5} b_2 \delta \psi_{123333} \,, \ \ 
\delta \psi_{111233} = b_1 \psi_{123333}\,, \nonumber\\ 
\delta \psi_{111113} =  - \frac{\sqrt{5}}{3} b_1 \delta \psi_{111333} \,, \ \ 
\delta \psi_{133333} = \frac{\sqrt{5}}{3} b_1 \psi_{111333}\,, \nonumber\\ 
\delta \psi_{112333} =  b_2 \delta \psi_{111123} \,, \ \ 
\delta \psi_{233333} = - \sqrt{5} b_1 \psi_{111123}\,,
\end{eqnarray}
leaving also three unconstrained parameters $\delta \psi_{111123}$, $\delta \psi_{111333}$ and $\delta \psi_{123333}$. The unconstrained small parameters $\delta \psi$ can be rotated away if we consider a $SO(3)$ basis transformation, 
$g_{\{\alpha^a \}} = \mathbb{1}_{3 \times 3} + \alpha^a \tau^a$
with $\alpha^{1}\to -{4 \sqrt{\frac{15}{7}} \delta \psi_{123333}}/{\mu_\psi }$, $\alpha^{2}\to {4 \sqrt{\frac{5}{21}} \delta \psi_{111333}}/{\mu_\psi }$, and $\alpha^{3}\to -{4 \sqrt{\frac{15}{7}} \delta \psi_{111123}}/{\mu_\psi }$. Therefore, Eqs.~\eqref{eq:w_singlet_psi} and \eqref{eq:w_nonet_psi} uniquely break $SO(3)$ to $A_5$. 

\subsection{Representation decomposition}  

After $SO(3)$ is broken to a non-Abelian discrete group, it is necessary to decompose each irrep of $SO(3)$ to a couple of irreps of the discrete one. This task is achieved by comparing reduction of Kronecker products of representations of $SO(3)$ with those of the discrete one \cite{Luhn:2011ip}. 

For irreps of $SO(3)$ decomposed to irreps of $A_4$, we identify $\underline{1}$, $\underline{3}$ of $SO(3)$ with $\mathbf{1}$, $\mathbf{3}$ of $A_4$, respectively and compare the Kronecker products 
\begin{eqnarray}
\underline{3} \times \underline{3} = \underline{1} + \underline{3} + \underline{5}
\end{eqnarray} 
in $SO(3)$ with 
\begin{eqnarray}
\mathbf{3} \times \mathbf{3} = \mathbf{1} + \mathbf{1}' + \mathbf{1}'' + \mathbf{3} + \mathbf{3}
\end{eqnarray}
in $A_4$. Since the right hand sides of both equations are identical, $\underline{5}$ of $SO(3)$ is decomposed to $\mathbf{1}' + \mathbf{1}'' + \mathbf{3}$ of $A_4$. One further compares right hand side of 
\begin{eqnarray}
\underline{3} \times \underline{5} = \underline{3} + \underline{5} + \underline{7}
\end{eqnarray} 
with that of 
\begin{eqnarray}
\mathbf{3} \times (\mathbf{1}' + \mathbf{1}'' + \mathbf{3}) = \mathbf{3}  + \mathbf{3} + \mathbf{1} + \mathbf{1}' + \mathbf{1}'' + \mathbf{3} + \mathbf{3}
\end{eqnarray} 
and obtains $\underline{7} = \mathbf{1} + \mathbf{3} + \mathbf{3}$, where $\mathbf{1}' \times \mathbf{3} = \mathbf{3}$ and $ \mathbf{1}'' \times \mathbf{3} = \mathbf{3}$ are used. Continuing to play this game, we can get decomposition of as high irrep of $SO(3)$ as we want into irreps of $A_4$. 

This game is directly applied into irrep decomposition in $S_4$ and $A_5$. In $S_4$, there are five irreps: $\mathbf{1}$ (the trivial singlet), $\mathbf{1}'$ (different from $\mathbf{1}'$ of $A_4$), $\mathbf{2}$, $\mathbf{3}$ and $\mathbf{3}'$. In $A_5$, there are five irreps: $\mathbf{1}$ (the trivial singlet), $\mathbf{3}$, $\mathbf{3'}$, $\mathbf{4}$ and $\mathbf{5}$. Keeping in mind the Kronecker products 
\begin{eqnarray}
&\mathbf{1}' \times \mathbf{1}' = \mathbf{1} \,, \ \ 
\mathbf{1}' \times \mathbf{2} = \mathbf{2} \,, \ \
\mathbf{2} \times \mathbf{2} = \mathbf{1} + \mathbf{1}' + \mathbf{2} \,, \nonumber\\
&\mathbf{3} \times \mathbf{3} = \mathbf{3}' \times \mathbf{3}' = \mathbf{1} + \mathbf{2} + \mathbf{3} + \mathbf{3}' \,, \ \
\mathbf{3} \times \mathbf{3}' = \mathbf{1}' + \mathbf{2} + \mathbf{3} + \mathbf{3}' 
\end{eqnarray}
in $S_4$, and 
\begin{eqnarray}
&\mathbf{3} \times \mathbf{3} = \mathbf{1} + \mathbf{3} + \mathbf{5} \,, \ \
\mathbf{3}' \times \mathbf{3}' = \mathbf{1} + \mathbf{3}' + \mathbf{5} \,, \ \
\mathbf{3} \times \mathbf{3}' = \mathbf{4} + \mathbf{5} \,, \nonumber\\
&\mathbf{3} \times \mathbf{4} = \mathbf{3}' + \mathbf{4} + \mathbf{5} \,, \ \
\mathbf{3}' \times \mathbf{4} = \mathbf{3} + \mathbf{4} + \mathbf{5} \,, \ \
\mathbf{3} \times \mathbf{5} = \mathbf{3}' \times \mathbf{5} = \mathbf{3} + \mathbf{3}' + \mathbf{4} + \mathbf{5} \,, \nonumber\\
&\mathbf{4} \times \mathbf{4} = \mathbf{1} + \mathbf{3} + \mathbf{3}' + \mathbf{4} + \mathbf{5} \,, \ \
\mathbf{4} \times \mathbf{5} = \mathbf{3} + \mathbf{3}' + \mathbf{3}' + \mathbf{4} + \mathbf{5} + \mathbf{5} \,, \nonumber\\
&\mathbf{5} \times \mathbf{5} = \mathbf{1} + \mathbf{3} + \mathbf{3}' + \mathbf{4} + \mathbf{4} + \mathbf{5} + \mathbf{5} 
\end{eqnarray}
in $A_5$, and comparing them with Kronecker products in $SO(3)$, we obtain irrep decompositions in $S_4$ and $A_5$, respectively. 

We summarise decomposition of irreps of $SO(3)$ (up to $\underline{13}$) to irreps of $A_4$, $S_4$ and $A_5$ in Table~\ref{tab:decomposition}. 
\begin{table}[h!]
\newcommand{\tabincell}[2]{\begin{tabular}{@{}#1@{}}#2\end{tabular}}
  \centering
  \begin{tabular}{ccccc}\hline\hline
  $SO(3)$ & $A_4$ & $S_4$ & $A_5$ \\\hline
  $\underline{1}$ & $\mathbf{1}$ & $\mathbf{1}$ & $\mathbf{1}$ \\
  $\underline{3}$ & $\mathbf{3}$ & $\mathbf{3}$ & $\mathbf{3}$ \\
  $\underline{5}$ & $\mathbf{1}' + \mathbf{1}'' + \mathbf{3}$ & $\mathbf{2} + \mathbf{3}'$ & $\mathbf{5}$ \\
  $\underline{7}$ & $\mathbf{1} + \mathbf{3} + \mathbf{3}$ & $\mathbf{1}' + \mathbf{3} + \mathbf{3}'$ & $\mathbf{3}' + \mathbf{4}$ \\
  $\underline{9}$ & $\mathbf{1} + \mathbf{1}' + \mathbf{1}'' + \mathbf{3} + \mathbf{3}$ & $\mathbf{1} + \mathbf{2} + \mathbf{3} + \mathbf{3}'$ & $\mathbf{4} + \mathbf{5}$ \\
  $\underline{11}$ & $\mathbf{1}' + \mathbf{1}'' + \mathbf{3} + \mathbf{3} + \mathbf{3}$ & $\mathbf{2} + \mathbf{3} + \mathbf{3} + \mathbf{3}'$ & $\mathbf{3} + \mathbf{3}' + \mathbf{5}$ \\
  $\underline{13}$ & $\mathbf{1} + \mathbf{1} + \mathbf{1}' + \mathbf{1}'' + \mathbf{3} + \mathbf{3} + \mathbf{3}$ & $\mathbf{1} + \mathbf{1}' + \mathbf{2} + \mathbf{3} + \mathbf{3}' + \mathbf{3}'$ & $\mathbf{1} + \mathbf{3} + \mathbf{4} + \mathbf{5}$ \\\hline\hline
\end{tabular}
  \caption{\label{tab:decomposition} Decomposition of some irreps of $SO(3)$ into irreps of $A_4$, $S_4$ and $A_5$. Results of decomposition to irreps of $A_4$ have been given in \cite{Berger:2009tt}. }
\end{table}

Before ending this section, we show more details of how a irrep of $SO(3)$ is decomposed into irreps of $A_4$ as follows, which will be useful for our discussion in the next two sections. 
\begin{itemize}

\item
For a triplet $\underline{3}$ of $SO(3)$, $\varphi = (\varphi_1, \varphi_2, \varphi_3)^T$, 
it is also a triplet $\mathbf{3}$ of $A_4$. 

\item
A $\underline{5}$-plet of $SO(3)$, $\chi$, can be represented as a rank-2 tensor $\chi_{ij}$ in the 3d space. It is symmetric, $\chi_{ij}=\chi_{ji}$, and traceless, $\chi_{11} + \chi_{22} + \chi_{33}=0$. Independent components can be chosen as $\chi_{11}$, $\chi_{12}$, $\chi_{13}$, $\chi_{23}$ and $\chi_{33}$. The $\underline{5}$-plet is decomposed to two non-trivial singlets $\mathbf{1}'$ and $\mathbf{1}''$ and one triplet $\mathbf{3}$ of $A_4$. It is useful to re-parametrise $\chi$ in the form
\begin{eqnarray}
\chi = 
\left(
\begin{array}{ccc}
 \frac{1}{\sqrt{3}}(\chi^{\prime}+\chi^{\prime\prime}) & \frac{1}{\sqrt{2}}\chi_3 & \frac{1}{\sqrt{2}}\chi_2 \\
 \frac{1}{\sqrt{2}}\chi_3 & \frac{1}{\sqrt{3}} (\omega \chi^{\prime}+ \omega^2 \chi^{\prime\prime}) & \frac{1}{\sqrt{2}}\chi_1 \\
 \frac{1}{\sqrt{2}}\chi_2 & \frac{1}{\sqrt{2}}\chi_1 & \frac{1}{\sqrt{3}}(\omega^2 \chi^{\prime}+ \omega \chi^{\prime\prime}) \\
\end{array}
\right)\,, \label{eq:rep_chi}
\end{eqnarray}
where $\omega = e^{2i\pi/3}$. This parametrisation has two advantages. One is the simple transformation property in $A_4$, 
\begin{eqnarray}
\chi^{\prime} \sim \mathbf{1}'\,,\quad
\chi^{\prime\prime} \sim \mathbf{1}''\,,\quad
\chi_{\mathbf{3}} \equiv (\chi_1,\chi_2,\chi_3) \sim \mathbf{3}  \text{ of } A_4\,.
\end{eqnarray}
The other is the normalised kinetic term, 
\begin{eqnarray}
\hspace{-8mm}
(\partial_\mu \chi^* \partial^\mu \chi)_{\underline{1}} &=& 
\partial_\mu \chi^{\prime*} \partial^\mu \chi' + 
\partial_\mu \chi^{\prime\prime*} \partial^\mu \chi'' +
\partial_\mu \chi^\dag_{\mathbf{3}} \partial^\mu \chi_{\mathbf{3}} \nonumber\\
&=&
\partial_\mu \chi^{\prime*} \partial^\mu \chi' + 
\partial_\mu \chi^{\prime\prime*} \partial^\mu \chi'' + 
\partial_\mu \chi^*_1 \partial^\mu \chi_1 + 
\partial_\mu \chi^*_2 \partial^\mu \chi_2 + 
\partial_\mu \chi^*_3 \partial^\mu \chi_3 \,.
\end{eqnarray}

\item
The $\underline{7}$-plet of $SO(3)$ is a symmetric and traceless rank-3 tensor in the 3d space. It is decomposed to one trivial singlet $\mathbf{1}$ and two triplets $\mathbf{3}$ of $A_4$. The former mentioned $\xi$ can be re-labelled as 
\begin{eqnarray}
&& \xi_{123}= \frac{1}{\sqrt{6}} \xi_{0} \,,\nonumber\\
&& \xi_{111}= -\frac{2}{\sqrt{10}} \xi'_{1}\,,\quad
\xi_{112}= \frac{1}{\sqrt{10}} \xi'_{2}-\frac{1}{\sqrt{6}} \xi_{2}\,,\quad
\xi_{113}= \frac{1}{\sqrt{10}} \xi'_{3}+\frac{1}{\sqrt{6}} \xi_{3}\,,\quad \nonumber\\
&& \xi_{133}= \frac{1}{\sqrt{10}} \xi'_{1}-\frac{1}{\sqrt{6}} \xi_{1}\,,\quad
\xi_{233}= \frac{1}{\sqrt{10}} \xi'_{2}+\frac{1}{\sqrt{6}} \xi_{2}\,,\quad
\xi_{333}= -\frac{2}{\sqrt{10}} \xi'_{3}\,. 
\end{eqnarray} 
Here, 
\begin{eqnarray}
\xi_0 \sim \mathbf{1}\,,\quad
\xi_{\mathbf{3}} \equiv (\xi_1,\xi_2,\xi_3) \sim \mathbf{3}\,, \quad 
\xi'_{\mathbf{3}} \equiv (\xi'_1,\xi'_2,\xi'_3) \sim \mathbf{3} \text{ of } A_4\,.
\end{eqnarray}
And the kinetic term is also normalised \footnote{Here we ignore the gauge interactions. Consequence of the gauge interactions will be given later in section~\ref{sec:gauge_interactions}}, 
\begin{eqnarray}
&&\hspace{-5mm}
(\partial_\mu \xi^* \partial^\mu \xi)_{\underline{1}} =
\partial_\mu \xi^*_0 \partial^\mu \xi_0 + 
\partial_\mu \xi^\dag_{\mathbf{3}} \partial^\mu \xi_{\mathbf{3}} +
\partial_\mu \xi^{\prime \dag}_{\mathbf{3}} \partial^\mu \xi^{\prime}_{\mathbf{3}}
\nonumber \\
&&\hspace{-1cm}= 
\partial_\mu \xi^*_0 \partial^\mu \xi_0 + 
\partial_\mu \xi^*_1 \partial^\mu \xi_1 + 
\partial_\mu \xi^*_2 \partial^\mu \xi_2 + 
\partial_\mu \xi^*_3 \partial^\mu \xi_3 +
\partial_\mu \xi^{\prime*}_1 \partial^\mu \xi^{\prime}_1 + 
\partial_\mu \xi^{\prime*}_2 \partial^\mu \xi^{\prime}_2 + 
\partial_\mu \xi^{\prime*}_3 \partial^\mu \xi^{\prime}_3 \,. \nonumber\\
\label{eq:kinetic_xi}
\end{eqnarray}
Since $\xi_0$ is a trivial singlet of $A_4$, once $\xi_0$ gets a non-zero VEV, $SO(3)$ will be broken but $A_4$ is still preserved. This is consistent with the discussion in the former subsection. 
\end{itemize}

\section{The further breaking of $A_4$ to residual $Z_3$ and $Z_2$ \label{sec:A4}}

In $A_4$ lepton flavour models, $A_4$ has to be broken to generate flavour mixing. In most of these models, residual symmetries $Z_3$ and $Z_2$ are preserved respectively in the charged lepton sector and neutrino sector after $A_4$ breaking. These residual symmetries are not precise but good approximate symmetries. The misalignment between $Z_3$ and $Z_2$ leading to a mixing with tri-bimaximal mixing pattern at leading order. 

Embedding $A_4$ to the continuous $SO(3)$ symmetry forces strong constraints on couplings, and the breaking of $A_4$ to $Z_3$ and $Z_2$ becomes very non-trivial. 
In this section, we will show, for definiteness, how to realise $A_4 \to Z_3$ and $Z_2$ in the framework of supersymmetric $SO(3)$-invariant theory. 
\subsection{$A_4 \to Z_3$}
The breaking of $A_4$ to $Z_3$ can be simply realised by using a triplet $\underline{3}$ of $SO(3)$. We denote such a flavon as $\varphi$. 
In order to obtain the $Z_3$-invariant VEV, we introduce an $\underline{1}$-plet driving field $\varphi^d_{\underline{1}}$ and a $\underline{5}$-plet driving field $\varphi^d_{\underline{5}}$ and consider the following $SO(3)$-invariant superpotential 
\begin{eqnarray}
w_\varphi = \varphi^d_{\underline{1}} \left( f_1 (\varphi \varphi)_{\underline{1}}-\mu_\varphi^2 \right) 
+ \frac{f_2}{\Lambda} \left( \varphi^d_{\underline{5}} \left(\xi (\varphi \varphi)_{\underline{5}} \right)_{\underline{5}} \right)_{\underline{1}} \,.
\end{eqnarray} 
Here as appearing in the non-renormalisable term, the scale $\Lambda$ is
assumed to be higher than the scale of $SO(3)$ breaking to $A_4$. 

Minimisation of the superpotential gives rise to 
\begin{eqnarray}
\frac{\partial w_\varphi}{\partial \varphi^d_{\underline{1}}} &=& f_1 (\varphi \varphi)_{\underline{1}} - \mu_\varphi^2= 0 \,, \nonumber\\
\frac{\partial w_\varphi}{\partial \varphi^d_{\underline{5}}} &=& \frac{f_2}{\Lambda} \left(\xi (\varphi \varphi)_{\underline{5}} \right)_{\underline{5}} = 0 \,,
\label{eq:w_phi}
\end{eqnarray} 
whose detailed formula is listed in Appendix~\ref{app:minimisation}. 
Starting from the $A_4$-invariant VEV $\langle \xi \rangle$ in Eq.~\eqref{eq:xi_VEV}, we use $\left(\xi (\varphi \varphi)_{\underline{5}} \right)_{\underline{5}} = 0$ to derive $\varphi_1^2 = \varphi_2^2 = \varphi_3^2$, and  $f_1 (\varphi \varphi)_{\underline{1}} - \mu_\varphi^2= 0$ to determine the value of $\varphi_1^2$. 
Here, we directly write out the following complete list of solutions
\begin{eqnarray}
\left(
\begin{array}{c}
 \langle \varphi_1 \rangle \\
 \langle \varphi_2 \rangle \\
 \langle \varphi_3 \rangle \\
\end{array}
\right) = \pm v_\varphi \left\{
\left(
\begin{array}{c}
 1 \\
 1 \\
 1 \\
\end{array}
\right)\,, \ \  
\left(
\begin{array}{c}
 1 \\
 -1 \\
 -1 \\
\end{array}
\right)\,, \ \  
\left(
\begin{array}{c}
 -1 \\
 1 \\
 -1 \\
\end{array}
\right)\,, \ \  
\left(
\begin{array}{c}
 -1 \\
 -1 \\
 1 \\
\end{array}
\right)
\right\} \,,
\label{eq:phi_VEV}
\end{eqnarray} 
where $v_\varphi = \mu_\varphi/\sqrt{3 f_1}$. 
For non-zero $v_\varphi$, all four VEVs break the $A_4$ symmetry. Each VEV preserves a different $Z_3$ group. In detail, $(1,1,1)^T$ preserves $Z_3^t = \{\mathbf{1}, t, t^2\}$, $(1,-1,-1)^T$ preserves $Z_3^{sts} = \{\mathbf{1}, sts, (sts)^2\}$, $(-1,1,-1)^T$ preserves $Z_3^{st} = \{\mathbf{1}, st, (st)^2\}$, and $(-1,-1,1)^T$ preserves $Z_3^{ts} = \{\mathbf{1}, ts, (ts)^2\}$. 
These $Z_3$ groups are conjugate to each each and have no physical difference \cite{Pascoli:2016eld, Morozumi:2017rrg}. 

Eq.~\eqref{eq:w_phi} involves interactions between $\varphi$ and $\xi$, specifically the non-renormalisable term which results in the breaking of $A_4$. These terms may influence the VEV of $\xi$ and shift it away from the $A_4$-invariant one. In general, this shifting effect is small enough due to suppression of the higher dimensional operator. In section \ref{sec:model}, we will construct a flavour model, and based on the model, we will discuss in detail the shift of the $\xi$ VEV due to non-normalisable interactions with the other flavons in section \ref{sec:subleadingvac}. As we will prove therein, the shift effect is suppressed by the scale $\Lambda$ and in general very small.

\subsection{$A_4 \to Z_2$ \label{sec:A4toZ2}}

We use the $\underline{5}$-plet $\chi$ to achieve the $A_4 \to Z_2$ breaking. The relevant superpotential terms could be considered as follows
\begin{eqnarray}
w_{\chi}=\chi_{\underline{1}}^d\left(g_1 (\chi \chi)_{\underline{1}} - \mu _{\chi }^2\right) + \frac{g_2}{\Lambda} \big( \chi_{\underline{3}}^d \left( \xi (\chi \chi)_{\underline{5}} \right)_{\underline{3}} \big)_{\underline{1}} + \frac{g_3}{\Lambda} \big( \chi_{\underline{3}}^d \left( \xi (\chi \chi)_{\underline{9}} \right)_{\underline{3}} \big)_{\underline{1}} + g_4 \big( \chi_{\underline{5}}^d (\chi \xi)_{\underline{5}} \big)_{\underline{1}} \,,
\end{eqnarray} 
where the driving fields $\chi^d_{\underline{1}}$, $\chi^d_{\underline{3}}$ and $\chi^d_{\underline{5}}$ are $\underline{1}$-, $\underline{3}$- and $\underline{5}$-plets of $SO(3)$. 
Minimisation of the superpotential results in equations
\begin{eqnarray}
\frac{\partial w_{\chi}}{\partial \chi_{\underline{1}}^d} &=& g_1 (\chi \chi)_{\underline{1}} - \mu _{\chi }^2 = 0 \,, \nonumber\\
\frac{\partial w_{\chi}}{\partial \chi_{\underline{3}}^d} &=& \frac{g_2}{\Lambda} \left( \xi (\chi \chi)_{\underline{5}} \right)_{\underline{3}} - \frac{g_3}{\Lambda} \left( \xi (\chi \chi)_{\underline{9}} \right)_{\underline{3}} = 0 \,, \nonumber\\
\frac{\partial w_{\chi}}{\partial \chi_{\underline{5}}^d} &=& g_4 (\chi \xi)_{\underline{5}} = 0 \,. \label{eq:w_chi}
\end{eqnarray} 
Given the $A_4$-invariant VEV $\langle \xi \rangle$ in Eq.~\eqref{eq:xi_VEV} as input, $(\chi \xi)_{\underline{5}} = 0$ leads to $\chi' = \chi'' = 0$. Then, $\left( \xi (\chi \chi)_{\underline{5}} \right)_{\underline{3}}$ takes the same form as $\left( \xi (\chi \chi)_{\underline{9}} \right)_{\underline{3}}$ and the requirement  $\left( \xi (\chi \chi)_{\underline{5}} \right)_{\underline{3}}=0 $ or $\left( \xi (\chi \chi)_{\underline{9}} \right)_{\underline{3}} = 0$ results in $\chi_1 \chi_2 = \chi_2 \chi_3 = \chi_3 \chi_1 = 0$. Therefore, two of $\chi_1$, $\chi_2$ and $\chi_3$ have to be zero. And the rest non-vanishing one is determined by $g_1 (\chi \chi)_{\underline{1}} - \mu _{\chi }^2 = 0$. We obtain the following complete list of solutions, 
\begin{eqnarray}
\left(
\begin{array}{c}
\langle \chi' \rangle \\
\langle \chi'' \rangle \\
\left(
\begin{array}{c}
 \langle \chi_1 \rangle \\
 \langle \chi_2 \rangle \\
 \langle \chi_3 \rangle \\
\end{array}
\right) 
\end{array}
\right)
= \left\{
\left(
\begin{array}{c}
0 \\
0 \\
\left(
\begin{array}{c}
 \pm v_{\chi} \\
 0 \\
 0 \\
\end{array}
\right)
\end{array}
\right)\,, \ \ 
\left(
\begin{array}{c}
0 \\
0 \\
\left(
\begin{array}{c}
 0 \\
 \pm v_{\chi} \\
 0 \\
\end{array}
\right)
\end{array}
\right)\,, \ \ 
\left(
\begin{array}{c}
0 \\
0 \\
\left(
\begin{array}{c}
 0 \\
 0 \\
 \pm v_{\chi} \\
\end{array}
\right)
\end{array}
\right)
\right\} \,. \label{eq:chi_VEV}
\end{eqnarray} 
where $v_\chi = \mu_\chi/\sqrt{g_1}$. 
These VEVs satisfy $Z_2$ symmetries. In details, the first, second, and third pairs preserve $Z_2^s=\{\mathbf{1}, s\}$, $Z_2^{tst^2}=\{\mathbf{1}, t s t^2\}$, $Z_2^{t^2st}=\{\mathbf{1}, t^2 s t\}$, respectively. All these VEVs are conjugate with each other and have no physical differences  \cite{Pascoli:2016eld, Morozumi:2017rrg}. There is a new scale $\mu_\chi$ introduced in the superpotential.

\subsection{Spontaneously splitting $\mathbf{1}'$ with $\mathbf{1}''$ of $A_4$}

In $A_4$ models, the three singlet irreps $\mathbf{1}$, $\mathbf{1}'$ and $\mathbf{1}''$ are usually assigned to $e^c$, $\mu^c$ and $\tau^c$ (or their permutation), respectively. These irreps are independent with each other in $A_4$ and the generated $e$, $\mu$ and $\tau$ masses are independent with each other. 

In the framework of $SO(3)$, the non-trivial singlet irreps $\mathbf{1}'$ and $\mathbf{1}''$ are obtained from the decomposition of $\underline{5}$ of $SO(3)$ (or higher irreps, e.g., $\underline{9}$ etc), as shown in Table~\ref{tab:decomposition}. These singlets are always correlated with each other. As a consequence, if we directly arrange two of the charged leptons (e.g., $\mu^c$ and $\tau^c$) to the same $\underline{5}$ of $SO(3)$, we have to face a fine tuning of masses of these two charged leptons.
In this subsection, we are going to consider how to avoid this problem from the spontaneous symmetry breaking of $A_4$.

We introduce another $\underline{5}$-plet flavon $\zeta$, 
\begin{eqnarray}
\zeta = 
\left(
\begin{array}{ccc}
 \frac{1}{\sqrt{3}}(\zeta^{\prime}+\zeta^{\prime\prime}) & \frac{1}{\sqrt{2}}\zeta_3 & \frac{1}{\sqrt{2}}\zeta_2 \\
 \frac{1}{\sqrt{2}}\zeta_3 & \frac{1}{\sqrt{3}} (\omega \zeta^{\prime}+ \omega^2 \zeta^{\prime\prime}) & \frac{1}{\sqrt{2}}\zeta_1 \\
 \frac{1}{\sqrt{2}}\zeta_2 & \frac{1}{\sqrt{2}}\zeta_1 & \frac{1}{\sqrt{3}}(\omega^2 \zeta^{\prime}+ \omega \zeta^{\prime\prime}) \\
\end{array}
\right)\,, \label{eq:rep_varphi}
\end{eqnarray}
and three driving fields $\zeta^d_{\underline{1}}$, $\zeta^d_{\underline{3}}$ and  $\tilde{\zeta}^d_{\underline{1}}$ with the following superpotential
\begin{eqnarray}
w_\zeta = \zeta^d_{\underline{1}} \left( \frac{h_1}{\Lambda} \left(\zeta (\zeta \zeta)_{\underline{5}} \right)_{\underline{1}} - \mu_\zeta^2 \right)
+ h_2 \big( \zeta^d_{\underline{3}} (\zeta \xi)_{\underline{3}} \big)_{\underline{1}} + h_3 \tilde{\zeta}^d_{\underline{1}} (\zeta \zeta)_{\underline{1}} \,. 
\end{eqnarray}
Minimisaiton of the superpotential gives to 
\begin{eqnarray}
\frac{\partial w_{\zeta}}{\partial \zeta_{\underline{1}}^d} &=& \frac{h_1}{\Lambda} \left( \zeta (\zeta \zeta)_{\underline{5}} \right)_{\underline{1}} - \mu_\zeta^2 = 0 \nonumber\\ 
\frac{\partial w_{\zeta}}{\partial \zeta_{\underline{3}}^d} &=& h_2 (\zeta \xi)_{\underline{3}} = 0 \nonumber\\
\frac{\partial w_{\zeta}}{\partial \tilde{\zeta}_{\underline{1}}^d} &=& h_3 ( \zeta \zeta )_{\underline{1}} = 0 \,.
\label{eq:minimisation_zeta}
\end{eqnarray}
The second row directly determines $\zeta_1 = \zeta_2 = \zeta_3 = 0$. It leaves the third row simplified to $( \zeta \zeta )_{\underline{1}} = 2 \zeta' \zeta'' = 0$, resulting in $\zeta^{\prime\prime} = 0$ (or $\zeta^{\prime} = 0$). The rest one, $\zeta'$ (or $\zeta''$), is determined by the first row, which is simplified to $\frac{h_1}{\Lambda} (\zeta')^3 - \mu_\zeta^2 = 0$, (or $\frac{h_1}{\Lambda} (\zeta'')^3 - \mu_\zeta^2 = 0$). These results are summarised as
\begin{eqnarray}
\left(
\begin{array}{c}
\langle \zeta' \rangle \\
\langle \zeta'' \rangle \\
\left(
\begin{array}{c}
 \langle \zeta_1 \rangle \\
 \langle \zeta_2 \rangle \\
 \langle \zeta_3 \rangle \\
\end{array}
\right) 
\end{array}
\right)
= \left\{
\left(
\begin{array}{c}
v_{\zeta} \omega^i \\
0 \\
\left(
\begin{array}{c}
 0 \\
 0 \\
 0 \\
\end{array}
\right)
\end{array}
\right)\,, \ \ 
\left(
\begin{array}{c}
0 \\
v_{\zeta} \omega^i \\
\left(
\begin{array}{c}
 0 \\
 0 \\
 0 \\
\end{array}
\right)
\end{array}
\right)
\right\} \,. \label{eq:zeta_VEV}
\end{eqnarray} 
with $v_{\zeta} = \sqrt[3]{\sqrt{3} \mu_\zeta^2 \Lambda / (2 h_1)}$ and $i=0,1,2$. 
We will see how this VEV can separate $\mu$ and $\tau$ masses in the next section.

To summarise, we realise the breaking of $A_4$ to $Z_3$ and $Z_2$ and achieve to split $\mathbf{1}'$ with $\mathbf{1}''$ of $A_4$ based on $SO(3)$-invariant superpotential. The scales representing the breaking of $A_4$, $v_\varphi$, $v_\chi$ and $v_\zeta$, should be much lower than the scale of $SO(3)$ breaking $v_\xi$. This can be satisfied by treating $\mu_\varphi^2$, $\mu_\chi^2$ and $\mu_\zeta^2$ as effective descriptions from higher dimensional operators.  One may notice that there may exist some unnecessary interactions which are not written out but cannot be forbidden based on current field arrangements. 
A detailed discussion on how to forbid the unnecessary coupling will be given in the next section on the model building. Besides, the ways to realise $A_4 \to Z_3$, $A_4 \to Z_2$ and split $\mathbf{1}'$ from $\mathbf{1}''$ showing above are not the unique ways. One can introduce different irreps, combined with different driving fields to achieve them. This difference further leads to the difference of model building, which will not be discussed in this paper. 

\section{A supersymmetric $A_4$ model from $SO(3) \times U(1)$ \label{sec:model}}
\subsection{The model}
\begin{figure}[h!]
\centering
\includegraphics[scale=.5]{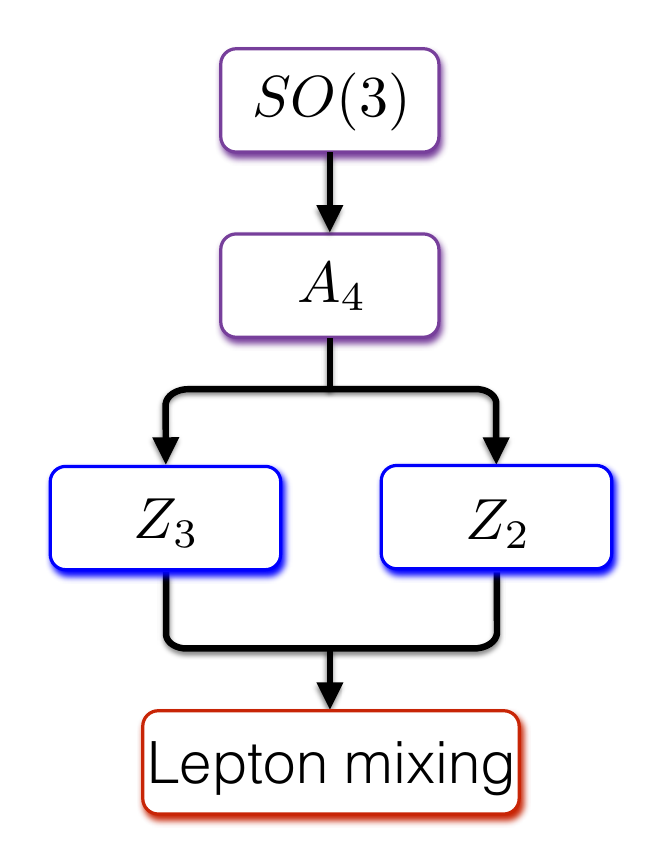}
\caption{\label{fig:sketch} A sketch of the symmetry breaking in the model and how flavour mixing is generated. The flavour symmetry at high energy is assumed to be supersymmetric $SO(3)$. It is broken first to $A_4$, which then breaks,
at a lower scale, to the residual symmetry $Z_3$ in the charged lepton sector and $Z_2$ in the neutrino sector,
with supersymmetry preserved throughout. The misalignment of the residual symmetries gives rise to flavour mixing. }
\end{figure}

In this section 
we will construct a supersymmetric $A_4$ model, based on $SO(3) \times U(1)$, with the breaking $SO(3) \to A_4$ and 
subsequently (at a lower scale) $A_4 \to Z_3, Z_2$, using the vacuum alignments discussed previously, where the misalignment of $Z_3$ in the charged lepton sector and $Z_2$ in the neutrino sector gives rise to lepton mixing. The model building strategy is shown in Fig.~\ref{fig:sketch}.  The $U(1)$ symmetry is used to forbid couplings which are unnecessary to generate the required flavon VEVs and flavour mixing. 
Note that no {\it ad hoc} discrete symmetries are introduced in this model. 

In $A_4$ models, the right-handed charged leptons $e^c$, $\mu^c$ and $\tau^c$ are arranged as $\mathbf{1}$, $\mathbf{1}'$ and $\mathbf{1}''$ (or their permutation), respectively. In $SO(3)$, the minimal irrep containing $\mathbf{1}'$ and $\mathbf{1}''$ is $\underline{5}$. In order to match with $A_4$ models, we embed $\mathbf{1}'$ and $\mathbf{1}''$ of $A_4$ to two different $\underline{5}$-plets of $SO(3)$. In our model, we embed $\mu^c$ and $\tau^c$ to two different $\underline{5}$-plets $R_\mu$ and $R_\tau$ \footnote{Imbedding $\mu^c$ and $\tau^c$ into the same $\underline{5}$-plet leads to fine tuning between $\mu$ and $\tau$ masses}. Four extra right-handed leptons are introduced for  $R_\mu$ and $R_\tau$, respectively. These particles should decouple at low energy theory to avoid unnecessary experimental constraints. We achieve this goal by introducing two left-handed $\underline{3}$-plets $L_\mu$, $L_\tau$ and two singlets $L_{\mu0}$, $L_{\tau0}$. 
We write out explicitly each components of the fermion multiplets in the 3d space as follows, 
\begin{eqnarray}
&&\ell = 
\begin{pmatrix}
 \ell_1 \\
 \ell_2 \\
 \ell_3 \\
\end{pmatrix} \,, \ \
N = 
\begin{pmatrix}
 N_1 \\
 N_2 \\
 N_3 \\
\end{pmatrix} \,, \ \
L_\mu = 
\begin{pmatrix}
 L_{\mu1} \\
 L_{\mu2} \\
 L_{\mu3} \\
\end{pmatrix} \,, \ \
L_\tau = 
\begin{pmatrix}
 L_{\tau1} \\
 L_{\tau2} \\
 L_{\tau3} \\
\end{pmatrix}\,, \ \
\nonumber\\
&& 
R_\mu = 
\begin{pmatrix}
 \frac{1}{\sqrt{3}}(\mu^c+R_\mu^{\prime\prime}) & \frac{1}{\sqrt{2}}R_{\mu3} & \frac{1}{\sqrt{2}}R_{\mu2} \\
 \frac{1}{\sqrt{2}}R_{\mu3} & \frac{1}{\sqrt{3}} (\omega \mu^c+ \omega^2 R_\mu^{\prime\prime}) & \frac{1}{\sqrt{2}}R_{\mu1} \\
 \frac{1}{\sqrt{2}}R_{\mu2} & \frac{1}{\sqrt{2}}R_{\mu1} & \frac{1}{\sqrt{3}}(\omega^2 \mu^c+ \omega R_\mu^{\prime\prime})
\end{pmatrix}\,,
 \nonumber\\
&&
R_\tau = 
\begin{pmatrix}
 \frac{1}{\sqrt{3}}(R_\tau^{\prime}+\tau^c) & \frac{1}{\sqrt{2}}R_{\tau3} & \frac{1}{\sqrt{2}}R_{\tau2} \\
 \frac{1}{\sqrt{2}}R_{\tau3} & \frac{1}{\sqrt{3}} (\omega R_\tau^{\prime}+ \omega^2 \tau^c) & \frac{1}{\sqrt{2}}R_{\tau1} \\
 \frac{1}{\sqrt{2}}R_{\tau2} & \frac{1}{\sqrt{2}}R_{\tau1} & \frac{1}{\sqrt{3}}(\omega^2 R_\tau^{\prime}+ \omega \tau^c)
\end{pmatrix} \,.
\label{eq:rep_fermion}
\end{eqnarray}
Here, $\ell_1 = (\nu_1,l_1)$, $\ell_2 = (\nu_2,l_2)$ and $\ell_3 = (\nu_3,l_3)$ are the three SM lepton doublets. $R_{\mu\mathbf{3}} \equiv (R_{\mu1}, R_{\mu2}, R_{\mu3})^T$ and $R_{\tau\mathbf{3}} \equiv (R_{\tau1}, R_{\tau2}, R_{\tau3})^T$ transform as $\mathbf{3}$ of $A_4$. 

Charges for all relevant fields in $SO(3) \times U(1)$ are listed in Table~\ref{tab:fields}. Besides $SO(3)$, we introduce additional $U(1)$ symmetry to forbid unnecessary couplings.


\begin{table}[h!]
\newcommand{\tabincell}[2]{\begin{tabular}{@{}#1@{}}#2\end{tabular}}
\centering
 \begin{tabular}{l c c c c c c c c c c c c c}\hline\hline\\[-4mm]
 Fields      & $\ell$                 & $N$                  & $e^c$             & $R_\mu$          & $R_\tau$           & $L_{\mu0}$           & $L_{\tau0}$     & $L_\mu$           & $L_\tau$        \\
 $SO(3)$ &  $\underline{3}$ & $\underline{3}$  & $\underline{5}$ & $\underline{5}$ & $\underline{1}$ & $\underline{1}$ & $\underline{1}$ & $\underline{3}$ & $\underline{3}$                                    \\
 $U(1)$    &  $-\frac{2}{3}$     & $+\frac{2}{3}$  & $-\frac{7}{3}$  &  $-1$                & $-\frac{1}{3}$     & $+\frac{5}{6}$         & $0$               & $+\frac{2}{3}$    & $0$  \\[2mm]\hline\\[-4mm]
 Fields     & $\eta$               & $\bar{\eta}$       & $\xi$                 & $\varphi$           & $\chi$               & $\zeta$              & $H_{u,d}$\\
 $SO(3)$ & $\underline{1}$ & $\underline{1}$ & $\underline{7}$ & $\underline{3}$ & $\underline{5}$ & $\underline{5}$ & $\underline{1}$                                  \\
 $U(1)$    & $+\frac{2}{3}$   & $-\frac{2}{3}$    & $+\frac{1}{3}$   & $+1$                   & $-\frac{4}{3}$   & $+\frac{1}{6}$   & $0$  \\[2mm]\hline\\[-4mm]
 Fields     & $\eta^d_{\underline{1}}$ & $\xi^d_{\underline{1}}$ & $\xi^d_{\underline{5}}$ & $\varphi^d_{\underline{1}}$ & $\varphi^d_{\underline{5}}$ & $\chi^d_{\underline{1}}$ & $\chi^d_{\underline{3}}$ & $\chi^d_{\underline{5}}$ &                                                             $\zeta^d_{\underline{1}}$ & $\zeta^d_{\underline{3}}$ & $\tilde{\zeta}^d_{\underline{1}}$                         \\
 $SO(3)$ & $\underline{1}$               & $\underline{1}$             & $\underline{5}$             & $\underline{5}$                   & $\underline{5}$                    & $\underline{1}$               & $\underline{3}$               & $\underline{5}$              & $\underline{1}$                 & $\underline{3}$                 & $\underline{1}$  \\
 $U(1)$   & $0$                                  & $-\frac{2}{3}$                & $-\frac{2}{3}$                & $-2$                                    & $-\frac{7}{3}$                        & $+2$                                & $+\frac{7}{3}$                             & $1$                 & $-\frac{1}{2}$                    & $-\frac{1}{2}$                    & $-\frac{1}{3}$ \\[2mm]\hline\hline
\end{tabular}  
  \caption{\label{tab:fields} Field arrangements in $SO(3) \times U(1)$ and decompositions of these fields in $A_4$ after $SO(3) \times U(1)$ is broken to $A_4$. }
\end{table}

\subsection{Vacuum alignments}  

Terms leading to $SO(3)$ breaking and $A_4$ breaking in the superpotential involving flavons and driving fields are given by 
\begin{eqnarray}
w_f &\supset& \eta^d_{\underline{1}} \left( d_1 \eta \bar{\eta} - \mu_\eta^2 \right) + \xi^d_{\underline{1}} \left( c_1 (\xi \xi)_{\underline{1}} - A_\xi \eta \right) + c_2 \big( \xi^d_{\underline{5}} (\xi \xi)_{\underline{5}} \big)_{\underline{1}} \nonumber\\ 
&& + \varphi _{\underline{1}}^d \left( f_1 (\varphi  \varphi )_{\underline{1}} - \frac{f_\varphi}{\Lambda} \eta^3 \right)
+ \frac{f_2}{\Lambda} \left( \varphi _{\underline{5}}^d \left( \xi (\varphi  \varphi )_{\underline{5}} \right)_{\underline{5}} \right)_{\underline{1}}  \nonumber\\
&& + \chi_{\underline{1}}^d \left( \frac{g_1'}{\Lambda} (\chi \chi)_{\underline{1}} \eta - \frac{g_\chi}{\Lambda} \bar{\eta}^3 
\right) 
+ \frac{g_2}{\Lambda} \big( \chi_{\underline{3}}^d \left( \xi (\chi \chi)_{\underline{5}} \right)_{\underline{3}} \big)_{\underline{1}} 
+ \frac{g_3}{\Lambda} \big( \chi_{\underline{3}}^d \left( \xi (\chi \chi)_{\underline{9}} \right)_{\underline{3}} \big)_{\underline{1}} 
+ g_4 \big( \chi_{\underline{5}}^d (\chi \xi)_{\underline{5}} \big)_{\underline{1}}  \nonumber\\
&& 
+ \tilde{\zeta}^d_{\underline{1}} \Big( \frac{h_1}{\Lambda}  \big( \zeta (\zeta \zeta)_{\underline{5}} \big)_{\underline{1}} 
- \frac{h_\zeta}{\Lambda^2} \left( \zeta (\varphi \chi)_{\underline{5}} \right)_{\underline{1}} \eta \Big)
+ h_2 \big( \zeta^d_{\underline{3}} (\zeta \xi)_{\underline{3}} \big)_{\underline{1}} 
+ h_3 \tilde{\zeta}^d_{\underline{1}} (\zeta \zeta)_{\underline{1}} + \cdots \,.
\label{eq:flavon_superpotential}
\end{eqnarray}
Here, the dots represent subleading corrections, which will be discussed in section~\ref{sec:subleadingvac}. Compared with the superpotential terms in sections~\ref{sec:SO3} and \ref{sec:A4}, Eq.~\eqref{eq:flavon_superpotential} takes a very similar form except the following differences:
\begin{itemize}
\item
The constant $\mu_\xi^2$ is not explicitly written out, but replaced by $A_\xi \eta$. Here $\eta$ and $\bar{\eta}$ are $SO(3)$ singlets. From the minimisation $\partial w_f / \partial \eta^d_{\underline{1}} = 0$, we know that both $\langle \eta \rangle$ and $\langle \bar{\eta} \rangle$ cannot be zero, and thus, we denote them as $v_\eta$ and $v_{\bar{\eta}}$, respectively. Once $\eta$ gets this VEV, $\mu_\xi^2 = A_\xi v_\eta$ is effectively obtained. This treatment is helpful for us to arrange charges for $\xi$. Otherwise only a $Z_2$ charge can be arranged for $\xi$. 
\item
The constants $\mu_\varphi^2$, $\mu_\chi^2$ and $\mu_\zeta^2$ are replaced by $f_\varphi \eta^3 / \Lambda$,  $g_\chi \bar{\eta}^3 / \Lambda$ and $h_\zeta \left( \zeta (\varphi \chi)_{\underline{5}} \right)_{\underline{1}} \eta / \Lambda^2$, respectively. These constants are just effective description of the higher dimensional operators after the relevant flavons get VEVs,
\begin{eqnarray}
\mu_\varphi^2 = \frac{f_\varphi}{\Lambda} v_{\eta}^3 \,, \ \ 
\mu_\chi^2 = \frac{g_\chi}{\Lambda} v_{\bar{\eta}}^3 \,, \ \ 
\mu_\zeta^2 = - i \sqrt{2} \frac{h_\zeta}{\Lambda^2} v_\zeta v_\varphi v_\chi v_\eta  \,.
\end{eqnarray}
\item
The term $g_1 \chi_{\underline{1}}^d (\chi \chi)_{\underline{1}}$ is not explicitly written out, but effectively obtained from the operator $ \frac{g_1'}{\Lambda} \chi_{\underline{1}}^d (\chi \chi)_{\underline{1}} \eta$ after $\eta$ gains the VEV. In this case, $g_1$ is effectively expressed as $g_1 = g_1' v_{\eta}/\Lambda$. The term $\left( \varphi _{\underline{5}}^d \big( \xi (\varphi  \varphi )_{\underline{5}} \big)_{\underline{5}} \right)_{\underline{1}}$ does not contribute since $(\varphi  \varphi )_{\underline{5}}$ vanishes at the $A_4$-invariant VEV. 
\end{itemize}

The approach for how the flavons obtained the required VEVs have been discussed in the former section. We do not repeat the relevant discussion here but just list the achieved VEVs of flavons, 
\begin{eqnarray}
\xi^{A_4}: && 
\langle \xi_{123} \rangle \equiv \frac{v_\xi}{\sqrt{6}} \,, \ \ \langle \xi_{111} \rangle = \langle \xi_{112} \rangle = \langle \xi_{113} \rangle = \langle \xi_{133} \rangle = \langle \xi_{233} \rangle = \langle \xi_{333} \rangle = 0 \,; \nonumber\\
\varphi^{Z_3}: && 
\left(
\begin{array}{c}
 \langle \varphi_1 \rangle \\
 \langle \varphi_2 \rangle \\
 \langle \varphi_3 \rangle \\
\end{array}
\right) = v_\varphi 
\left(
\begin{array}{c}
 1 \\
 1 \\
 1 \\
\end{array}
\right)\,; \nonumber\\
\chi^{Z_2}: && 
\left(
\begin{array}{c}
\langle \chi' \rangle \\
\langle \chi'' \rangle \\
\left(
\begin{array}{c}
 \langle \chi_1 \rangle \\
 \langle \chi_2 \rangle \\
 \langle \chi_3 \rangle \\
\end{array}
\right) 
\end{array}
\right)
= v_{\chi}
\left(
\begin{array}{c}
0 \\
0 \\
\left(
\begin{array}{c}
 1 \\
 0 \\
 0 \\
\end{array}
\right)
\end{array}
\right)\,; \nonumber\\
\zeta^{\mathbf{1}'}: &&
\left(
\begin{array}{c}
\langle \zeta' \rangle \\
\langle \zeta'' \rangle \\
\left(
\begin{array}{c}
 \langle \zeta_1 \rangle \\
 \langle \zeta_2 \rangle \\
 \langle \zeta_3 \rangle \\
\end{array}
\right) 
\end{array}
\right)
= v_{\zeta}
\left(
\begin{array}{c}
1 \\
0 \\
\left(
\begin{array}{c}
 0 \\
 0 \\
 0 \\
\end{array}
\right)
\end{array}
\right) \,.
\label{eq:flavon_VEV}
\end{eqnarray} 
where $v_\xi$, $v_\varphi$, $v_\chi$ and $v_\zeta$ are respectively given by 
\begin{eqnarray}
v_\xi = \sqrt{\frac{A_\xi v_\eta}{c_1}} \,, \ \
v_\varphi = v_{\eta} \sqrt{\frac{f_\varphi v_{\eta}}{3 f_1 \Lambda}}  \,, \ \ 
v_\chi = v_{\bar{\eta}} \sqrt{\frac{g_\chi v_{\bar{\eta}}}{g_1' v_{\eta}} } \,, \ \ 
v_\zeta = v_{\eta} \left( \frac{-f_\varphi g_\chi h_\zeta^2 v_{\bar{\eta}}^3}{2 f_1 g_1' h_1^2 \Lambda^3} \right)^{\frac{1}{4}}  \,.
\label{eq:VEV}
\end{eqnarray}

We briefly discuss the scales involved in the model. 
The VEV $v_\xi$ represents the scale of $SO(3) \to A_4$ and $v_\varphi$ and $v_\chi$ represent the scales of $A_4 \to Z_3$ and $Z_2$, respectively. VEVs of $\eta$ and $\bar{\eta}$ do not break any non-Abelian symmetries but $U(1)$, their role is to connect the scales of $SO(3)$ breaking and $A_4$ breakings. For the scale of $A_4 \to Z_3$, $v_\varphi \ll v_{\eta}$ is naturally achieved due to the suppression of $\Lambda$ in the dominator in Eq.~\eqref{eq:VEV}. For the scale of $A_4 \to Z_2$, $v_\chi \ll v_\eta$ can be achieved by either assuming a hierarchy $v_\eta \ll v_{\bar{\eta}}$ all assuming a small coefficient $g_\chi$. The VEV $v_\xi$ can be much larger than $v_\eta$ and $v_{\bar{\eta}}$ if the dimension one parameter $A_\xi$ is large enough. With the above treatment (but not the unique treatment), we can easily achieve a hierarchy of energy scales 
\begin{eqnarray}
\text{UV scale } (\Lambda) \gg \text{ scale of } SO(3) \to A_4 \ (v_\xi) ~ \gg ~ \text{ scales of } A_4 \to Z_3, Z_2 \ (v_\varphi, v_\chi) \,.
\label{eq:hierarchy}
\end{eqnarray}

In the following, we simplify our discussion by assuming all dimensionless parameters in the flavon superpotential being of order one. In this case, orders of magnitude of $v_\xi$, $v_\varphi$, $v_\chi$ and $v_\zeta$ are determined by $\Lambda$, $A_\xi$, $v_{\bar{\eta}}$ and $v_\eta$ as
\begin{eqnarray}
v_\xi \sim v_\eta \sqrt{\frac{A_\xi}{v_\eta}} \,, \ \ 
v_\varphi \sim v_{\eta} \sqrt{\frac{v_{\eta}}{\Lambda}} \,, \ \ 
v_\chi \sim v_{\bar{\eta}} \sqrt{ \frac{v_{\bar{\eta}}}{v_{\eta}} } \,, \ \ 
v_\zeta \sim v_{\eta} \left( \frac{v_{\bar{\eta}}}{\Lambda} \right)^{\frac{3}{4}}\,. \ \ 
\label{eq:VEV_maglitude}
\end{eqnarray}
The hierarchy in Eq.~\eqref{eq:hierarchy} is obtained by requiring $\Lambda \gg A_\xi \gg v_{\bar{\eta}} \gg v_{\eta}$. 


\subsection{Lepton masses}

Lagrangian terms for generating charged lepton masses are given by 
\begin{eqnarray}
w_\ell = w_{e^c} + w_{R_\mu} + w_{R_\tau} + w_{N} 
\end{eqnarray}
with
\begin{eqnarray}
w_{e^c} &\supset& \frac{y_{e1}}{\Lambda^3} (\varphi \varphi )_{\underline{1}} (\varphi \ell )_{\underline{1}} e^c H_d + 
\frac{y_{e2}}{\Lambda^3} \Big( \big( (\varphi \varphi )_{\underline{5}} \varphi)_{\underline{3}} \ell \Big)_{\underline{1}} e^c H_d\,, \nonumber\\
w_{R_\mu} &\supset& \frac{y_{\mu1}}{\Lambda^2}\big( \varphi (\ell R_{\mu})_{\underline{3}} \big)_{\underline{1}} \bar{\eta} H_d + 
\frac{y_{\mu2}}{\Lambda} \big( \varphi (L_{\mu} R_\mu)_{\underline{3}} \big)_{\underline{1}} \bar{\eta} + 
Y_{\mu1} L_{\mu0} \left(\zeta R_{\mu}\right)_{\underline{1}} \nonumber\\
&& + \frac{Y_{\mu3}}{\Lambda}\big( \xi (\ell R_{\mu})_{\underline{7}} \big)_{\underline{1}} H_d + 
Y_{\mu2} \big( \xi (L_{\mu} R_{\mu})_{\underline{5}} \big)_{\underline{1}} \,, \nonumber\\
w_{R_\tau} &\supset& \frac{y_{\tau}}{\Lambda} \big( \varphi (\ell R_\tau)_{\underline{3}} \big)_{\underline{1}} H_d + 
\frac{Y_{\tau1}}{\Lambda} L_{\tau0} \big( (\zeta \zeta)_{\underline{5}} R_\tau \big)_{\underline{1}}+Y_{\tau2} \big( \xi (L_{\tau} R_\tau)_{\underline{5}} \big)_{\underline{1}} \,, \nonumber\\
w_N &\supset& y_N (\ell N)_{\underline{1}} H_u + \frac{\lambda_\eta}{\Lambda} \bar{\eta}^2 (N N)_{\underline{1}} + \lambda_\chi \big( \chi (N N)_{\underline{5}} \big)_{\underline{1}}
\end{eqnarray}
at leading order. After the flavons get their VEVs, we arrive at the effective Yukawa couplings for leptons and Majorana masses for right-handed neutrinos. 

The Yukawa coupling of $e^c$ is given by
\begin{eqnarray}
w_e^{\text{eff}} = y_e \frac{v_\varphi^3}{\Lambda^3} \ \ell^T 
\begin{pmatrix}
 1\\
 1\\
 1\\
\end{pmatrix}
e^c H_d
\label{eq:e_yukawa}
\end{eqnarray}
where $y_e = 3 y_{e1} + 4 y_{e2}$. 

Couplings involving $R_\mu$ are given by
\begin{eqnarray}
w_{R_\mu}^{\text{eff}} = (\ell^T, L_{\mu0},L_{\mu\mathbf{3}}^T) \ 
\begin{pmatrix}
y_{\mu1} \frac{v_\varphi v_{\bar{\eta}}}{\sqrt{3}\Lambda^2} V_\omega H_d  & 
y_{\mu1} \frac{v_\varphi v_{\bar{\eta}}}{\sqrt{3}\Lambda^2} V_\omega^* H_d & 
2 \sqrt{3} Y_{\mu3} \frac{v_\xi}{\Lambda} \ \mathbb{1}_{3\times3} \ H_d \\
 0 & Y_{\mu1} v_\zeta & \mathbb{0}_{1\times3} \\
 y_{\mu2} \frac{v_\varphi v_{\bar{\eta}}}{\sqrt{3}\Lambda} V_\omega & y_{\mu2} \frac{v_\varphi v_{\bar{\eta}}}{\sqrt{3}\Lambda} V_\omega^* & 2 \sqrt{3} Y_{\mu2} v_\xi \ \mathbb{1}_{3\times3} \\
\end{pmatrix}
\begin{pmatrix}
 \mu^c\\
 R_\mu''\\
 R_{\mu\mathbf{3}}\\
\end{pmatrix} \,.
\end{eqnarray}
where 
\begin{eqnarray}
V_\omega = \begin{pmatrix} 1 \\ \omega \\ \omega^2 \end{pmatrix} \,, \ \ 
V_\omega^* = \begin{pmatrix} 1 \\ \omega^2 \\ \omega \end{pmatrix} \,.
\end{eqnarray}
$L_{\mu\mathbf{3}}$ and $R_{\mu\mathbf{3}}$ obtain three degenerate heavy masses $2 \sqrt{3} Y_{\mu2} v_\xi$. These mass are much heavier than the electroweak scale, and thus for the low energy theory, $L_{\mu\mathbf{3}}$ and $R_{\mu\mathbf{3}}$ decouple. $L_{\mu0}$ and $R_{\mu}''$ obtain a mass $Y_{\mu1} v_\zeta$. For $v_\zeta$ heavier than the electroweak scale, $R_{\mu}''$ decouples from the low energy theory. In this way, we successfully split $R_{\mu}''$ with $\mu^c$.
After the heavy leptons are integrated out, we are left with the following couplings at the low energy theory,  
\begin{eqnarray}
w_\mu^{\text{eff}} =y_\mu \frac{v_\varphi v_{\bar{\eta}}}{\sqrt{3}\Lambda^2}  \  \ell^T \ 
\begin{pmatrix} 1 \\ \omega \\ \omega^2 \end{pmatrix}
\mu^c H_d \,,
\label{eq:mu_yukawa}
\end{eqnarray}
where $y_\mu = y_{\mu1} - y_{\mu2} Y_{\mu3}/Y_{\mu2}$ with the $y_{\mu 2}$ term obtained via a seesaw-like formula. 

Those coupling to $R_\tau$ are given by
\begin{eqnarray}
w_{R_\tau}^{\text{eff}} =  (\ell^T, L_{\tau0}, L_{\tau\mathbf{3}}^T) \ 
\begin{pmatrix}
 y_\tau \frac{v_\varphi}{\sqrt{3}\Lambda} V_\omega^* H_d & 
 y_\tau \frac{v_\varphi}{\sqrt{3}\Lambda} V_\omega H_d & 
 \mathcal{O}(y_\tau \frac{v_\varphi}{\sqrt{3}\Lambda}) H_d \\
 0 & Y_{\tau1} \frac{2v_\zeta^2}{\sqrt{3} \Lambda} & \mathbb{0}_{1\times3} \\
 \mathbb{0}_{3\times1} & \mathbb{0}_{3\times1} & 2 \sqrt{3} Y_{\tau2} v_\xi \ \mathbb{1}_{3\times3} \\
\end{pmatrix}
\begin{pmatrix}
 \tau^c\\
 R_\tau'\\
 R_{\tau\mathbf{3}}\\
\end{pmatrix} \,.
\end{eqnarray}
$L_{\tau\mathbf{3}}$ and $R_{\tau\mathbf{3}}$ obtain three degenerate heavy masses $2 \sqrt{3} Y_{\tau2} v_\xi$, which are much heavier than the electroweak scale. $L_{\tau0}$ and $R_{\tau}'$ obtain a mass $2Y_{\tau1} v_\zeta^2/(\sqrt{3} \Lambda)$. This mass term should also be heavier than the electroweak scale such that $R_{\tau}'$ can decouple from the low energy theory. This mass term aims to split $R_{\tau}'$ with $\tau^c$ and it provides a stronger constraint to the scale $v_\zeta$ than that splitting $R_{\mu}''$ with $\mu^c$. After all these heavy particles decouple, we obtain Yukawa coupling for $\tau^c$ at low energy as
\begin{eqnarray}
w_\tau^{\text{eff}} = y_\tau \frac{v_\varphi}{\sqrt{3}\Lambda} \ \ell^T \ 
\begin{pmatrix} 1 \\ \omega^2 \\ \omega \end{pmatrix}
\tau^c H_d \,.
\label{eq:tau_yukawa}
\end{eqnarray}

After the Higgs $H_d$ gets the VEV $\langle H_d \rangle = v_d/\sqrt{2}$, we arrive at the charged lepton mass matrix
\begin{eqnarray}
M_l= \begin{pmatrix}
 y_e \frac{v_\varphi^3}{\Lambda^3} & y_\mu \frac{v_\varphi v_{\bar{\eta}}}{\sqrt{3} \Lambda^2} & y_\tau \frac{v_\varphi}{\sqrt{3} \Lambda} \\
 y_e \frac{v_\varphi^3}{\Lambda^3} & \omega y_\mu \frac{v_\varphi v_{\bar{\eta}}}{\sqrt{3} \Lambda^2} & \omega^2 y_\tau \frac{v_\varphi}{\sqrt{3} \Lambda} \\
 y_e \frac{v_\varphi^3}{\Lambda^3} & \omega^2 y_\mu \frac{v_\varphi v_{\bar{\eta}}}{\sqrt{3} \Lambda^2} & \omega y_\tau \frac{v_\varphi}{\sqrt{3} \Lambda} \\
\end{pmatrix} \frac{v_d}{\sqrt{2}}
\end{eqnarray}
in the basis $\ell$ and $(e^c,\mu^c,\tau^c)^T$. 
This matrix is diagonal by a unitary matrix $U_l$ via $U_l^T M_l = \text{diag} \{ m_e, m_\mu, m_\tau \}$ with 
\begin{eqnarray}
U_l = U_\omega \equiv \frac{1}{\sqrt{3}} \begin{pmatrix}
 1 & 1 & 1 \\
 1 & \omega^2 & \omega \\
 1 & \omega & \omega^2 \\
\end{pmatrix}
\end{eqnarray}
and 
\begin{eqnarray}
m_e = \left|\sqrt{3} y_e \frac{v_\varphi^3 v_d}{\sqrt{2} \Lambda^3} \right| \, \ \ 
m_\mu = \left| y_\mu \frac{v_\varphi v_{\bar{\eta}} v_d}{\sqrt{2} \Lambda^2} \right| \, \ \
m_\tau = \left| y_\tau \frac{v_\varphi v_d}{\sqrt{2}\Lambda} \right| \,. 
\end{eqnarray}
In this model, since $y_e$, $y_\mu$, and $y_\tau$ are totally independent free parameters, there is no need to introduce a fine tuning of them to fit the hierarchy of $e$, $\mu$ and $\tau$ masses. 

If we naturally assume the dimensionless parameters $y_e$, $y_\mu$ and $y_\tau$ are all of order one, we then obtain $m_e : m_\mu : m_\tau \sim (v_{\varphi} / \Lambda)^2 :v_{\bar{\eta}} / \Lambda : 1$. The mass between $L_{\tau0}$ and $R_{\tau}'$ is of order $v_\zeta^2/ \Lambda$. It should be much heavier than the electroweak scale to avoid the constraints from collider searches, i.e., $v_\zeta^2 / v_\varphi \gg v_d$.

The realisation of neutrino masses is straightforward. The relevant superpotential terms at leading order are given by
\begin{eqnarray}
w_N = y_N (\ell N)_{\underline{1}} H_u + \frac{\lambda_\eta}{\Lambda} \bar{\eta}^2 (N N)_{\underline{1}} + \lambda_\chi \big( \chi (N N)_{\underline{5}} \big)_{\underline{1}} \,.
\end{eqnarray}
The generated Dirac mass matrix between $\nu$ and $N$ and Majorana mass matrix for $N$, in the bases $(\nu_1,\nu_2,\nu_3)^T$ and $(N_1, N_2, N_3)^T$, are respectively given by 
\begin{eqnarray}
M_D &=& \frac{y_D v_u}{\sqrt{2}} \mathbb{1}_{3\times3} \,, \ \nonumber\\
M_M &=& \begin{pmatrix}
 a & 0 & 0 \\
 0 & a & b \\
 0 & b & a \\
\end{pmatrix} \,,
\end{eqnarray} 
where 
$a = 2\lambda_{\bar{\eta}} v_{\bar{\eta}}^2 /\Lambda$ and $b = 2 \sqrt{2} \lambda_\chi v_\chi$. 
It is straightforward to diagonalise $M_M$ via $U_\nu^\dag M_M U_\nu^* = \text{diag} \{ M_1, M_2, M_3 \}$ with 
\begin{eqnarray}
U_{\nu} = \begin{pmatrix}
 0 & 1 & 0 \\
 \frac{1}{\sqrt{2}} & 0 & \frac{i}{\sqrt{2}} \\
 \frac{1}{\sqrt{2}} & 0 & \frac{-i}{\sqrt{2}} \\
\end{pmatrix} P_\nu \,,
\end{eqnarray}
where $P_\nu = {\rm diag} \set{ e^{i \frac{\beta_1}{2}}, e^{i \frac{\beta_2}{2}}, e^{i \frac{\beta_3}{2}}}$ and
\begin{eqnarray}
M_1 = |a+b|\,, \ \ M_2 &=& |a|\,, \ \ M_3 = |a-b| \,, \\
\beta_1= \arg \left( b + a \right) \,, \ \
\beta_2 &=& \arg \left( a \right) \,, \ \
\beta_3= \arg \left( b - a \right) \,.
\end{eqnarray}
Applying the seesaw mechanism, $M_\nu = - M_D M_M^{-1} M_D^T$, we obtain that $M_\nu$ is diagonalised as $U_\nu^T M_\nu U_\nu = \text{diag} \set{ m_1, m_2, m_3 }$.  The three mass eigenvalues for light neutrinos are given by $m_{1} = y_D^2 v_u^2 / (2 |b+a|)$, $m_2 = y_D^2 v_u^2 / (2 |a|)$ and $m_{3} = y_D^2 v_u^2 / (2 |b-a|)$. 
The PMNS matrix is given by $U_{\rm PMNS} = U_l^\dag U_\nu = U_{\rm TBM} P_\nu$. We are left with the tri-bimaximal mixing.

In the model discussed so far, the crucial point in deriving an $A_4$-invariant VEV is the requirement $(\xi\xi)_{\underline{5}} = 0$, while those to derive a $Z_2$- or $Z_3$-invariant vacuum is the requirement $\left(\xi (\varphi \varphi)_{\underline{5}} \right)_{\underline{5}} = 0$ or $(\chi \xi)_{\underline{5}}= \left( \xi (\chi \chi)_{\underline{5}} \right)_{\underline{3}} = 0$, respectively. These requirements are obtained via the minimisation of the superpotential. However, extra terms may be involved in the superpotential and lead to that the above requirements do not hold explicitly. As a consequence, the relevant vacuums do not preserve the symmetries explicitly. In the next subsection, we will prove that after including these terms, the flavon VEVs do deviate from the former symmetric ones, but the size of the deviations are safely very small. Then in the subsequent subsection we consider subleading effects to the flavour mixing and show that it gives important corrections.

\subsection{Subleading corrections to the vacuum (are negligible) \label{sec:subleadingvac}}


\begin{table}[t!]
  \centering
  \begin{tabular}{p{3mm} p{24mm} p{28mm} p{87mm}}\hline\hline
 & \multirow{2}{25mm}{renormalisable terms $w_f^{d \leqslant 3}$} 
 & \multirow{2}{30mm}{quartic terms \\ $w_f^{d=4} \times \Lambda$} 
 & \multirow{2}{95mm}{quintic terms \\ $w_f^{d=5} \times \Lambda^2$} \\
 \\[1mm]\hline\\[-4mm]
 
 $\eta^d_{\underline{1}}$
 & \multirow{2}{24mm}{$\mu_\eta^2 \eta _{\underline{1}}^d$,\\ $\eta _{\underline{1}}^d \eta  \bar{\eta }$}
 & \multirow{2}{32mm}{\footnotesize
 $\eta _{\underline{1}}^d (\xi  \xi )_{\underline{1}} \bar{\eta }$, 
 $\eta _{\underline{1}}^d \big( (\varphi  \chi )_{\underline{7}} \xi \big)_{\underline{1}}$}
 & \multirow{1}{87mm}{\footnotesize
 $\eta _{\underline{1}}^d \eta ^2 \bar{\eta }^2$, 
 $\eta _{\underline{1}}^d \big( (\xi  \xi )_{\underline{5}} \chi \big)_{\underline{1}} \eta$, 
 $\eta _{\underline{1}}^d \big( (\varphi  \varphi )_{\underline{5}} \chi \big)_{\underline{1}} \bar{\eta}$, 
 $\eta _{\underline{1}}^d \big( (\zeta  \zeta )_{\underline{5}} (\varphi  \chi)_{\underline{5}}\big)_{\underline{1}}$} \\
 \\
   \\[-4mm]\hline\\[-5mm]
  
 $\xi^d_{\underline{1}}$
 & \multirow{2}{15mm}{$A_\xi \xi _{\underline{1}}^d \eta$, \\ $\xi _{\underline{1}}^d \left(\xi  \xi \right)_{\underline{1}}$}
 & \multirow{2}{24mm}{\footnotesize
 $\xi _{\underline{1}}^d \eta ^2 \bar{\eta }$, 
 $\eta _{\underline{1}}^d \big( (\varphi  \varphi )_{\underline{5}} \chi \big)_{\underline{1}}$
 }
 & \multirow{2}{90mm}{\footnotesize
 $\xi _{\underline{1}}^d (\xi  \xi )_{\underline{1}} \eta \bar{\eta }$, 
 $\xi _{\underline{1}}^d \big( (\varphi  \chi )_{\underline{7}} \xi \big)_{\underline{1}} \eta$, 
 $\xi _{\underline{1}}^d (\varphi  \varphi)_{\underline{1}} \bar{\eta}^2$, 
 $\xi _{\underline{1}}^d \big( (\zeta  \zeta)_{\underline{1}} \big)^2$, \\
 $\xi _{\underline{1}}^d \left( (\zeta  \zeta)_{\underline{5}} (\zeta  \zeta)_{\underline{5}}\right)_{\underline{1}}$,
 $\xi _{\underline{1}}^d \left( (\zeta  \zeta)_{\underline{9}} (\zeta  \zeta)_{\underline{9}}\right)_{\underline{1}}$ }
 \\
 \\
  \\[-4mm]\hline\\[-5mm]

 $\xi^d_{\underline{5}}$
 & $\left( \xi _{\underline{5}}^d (\xi  \xi)_{\underline{5}} \right)_{\underline{1}}$
 & \multirow{5}{35mm}{\footnotesize
 $\left( \xi _{\underline{5}}^d \left(\xi \varphi \right)_{\underline{5}} \right)_{\underline{1}} \bar{\eta}$, 
 $( \xi _{\underline{5}}^d \chi )_{\underline{1}} (\varphi  \varphi)_{\underline{1}}$, 
 $\left( \xi _{\underline{5}}^d \left( \xi (\zeta  \zeta)_{\underline{5}} \right)_{\underline{5}} \right)_{\underline{1}}$,
 $\left( \xi _{\underline{5}}^d \left( \xi (\zeta  \zeta)_{\underline{9}} \right)_{\underline{5}} \right)_{\underline{1}}$,
 $\left( \xi _{\underline{5}}^d \left( \chi (\varphi  \varphi)_{\underline{5}} \right)_{\underline{5}} \right)_{\underline{1}}$
 }
 & \multirow{4}{75mm}{\footnotesize
 $( \xi _{\underline{5}}^d \chi )_{\underline{1}} \eta ^3 $, 
 $\left( \xi _{\underline{5}}^d (\xi  \xi)_{\underline{5}} \right)_{\underline{1}} \eta \bar{\eta} $, 
 $\left( \xi _{\underline{5}}^d \big( \xi (\varphi \chi )_{\underline{3}} \big)_{\underline{5}} \right)_{\underline{1}} $,
 $\left( \xi _{\underline{5}}^d \big( \xi (\varphi \chi )_{\underline{5}} \big)_{\underline{5}} \right)_{\underline{1}} $,
 $\left( \xi _{\underline{5}}^d \big( \xi (\varphi \chi )_{\underline{7}} \big)_{\underline{5}} \right)_{\underline{1}} $,
 $\left( \xi _{\underline{5}}^d \left( \varphi \varphi \right)_{\underline{5}} \right)_{\underline{1}} \bar{\eta}^2$, 
 $\left( \xi _{\underline{5}}^d \big( \varphi (\zeta \zeta)_{\underline{5}} \big)_{\underline{5}} \right)_{\underline{1}} \bar{\eta}$, 
 $\left( \xi _{\underline{5}}^d (\zeta  \zeta)_{\underline{5}} \right)_{\underline{1}} (\zeta  \zeta)_{\underline{1}}$,
 $\left( \xi _{\underline{5}}^d \big( (\zeta  \zeta)_{\underline{5}} (\zeta  \zeta)_{\underline{5}} \big)_{\underline{5}} \right)_{\underline{1}}$,
 $\left( \xi _{\underline{5}}^d \big( (\zeta  \zeta)_{\underline{5}} (\zeta  \zeta)_{\underline{9}} \big)_{\underline{5}} \right)_{\underline{1}}$,
 $\left( \xi _{\underline{5}}^d \big( (\zeta  \zeta)_{\underline{9}} (\zeta  \zeta)_{\underline{9}} \big)_{\underline{5}} \right)_{\underline{1}}$
 } \\
 \\
 \\
 \\
 \\
 \\
 \\[1mm]\hline\\[-5mm]
  
 $\varphi^d_{\underline{1}}$
 & $\varphi _{\underline{1}}^d (\varphi  \varphi )_{\underline{1}}$
 & $\varphi _{\underline{1}}^d \eta^3$
 & \multirow{1}{80mm}{\footnotesize
 $\varphi _{\underline{1}}^d (\xi  \xi )_{\underline{1}} \eta^2$, 
 $\varphi _{\underline{1}}^d (\varphi  \varphi )_{\underline{1}} \eta  \bar{\eta }$,
 $\varphi _{\underline{1}}^d \left( (\xi  \xi )_{\underline{5}} (\xi  \varphi )_{\underline{5}} \right)_{\underline{1}}$, 
 $\varphi _{\underline{1}}^d \left( (\xi  \xi )_{\underline{9}} (\xi  \varphi )_{\underline{9}} \right)_{\underline{1}}$ 
 }\\
 \\
  \\[-4mm]\hline\\[-5mm]
  
 $\varphi^d_{\underline{5}}$
 & 0
 & $\left( \varphi _{\underline{5}}^d \left( \xi (\varphi  \varphi )_{\underline{5}} \right)_{\underline{5}} \right)_{\underline{1}}$
 & \multirow{2}{95mm}{\footnotesize
 $\left( \varphi _{\underline{5}}^d \big( (\xi \xi )_{\underline{5}} \varphi \big)_{\underline{5}}\right)_{\underline{1}} \eta$,
 $\left( \varphi _{\underline{5}}^d (\varphi \varphi )_{\underline{5}}\right)_{\underline{1}} (\zeta  \zeta )_{\underline{1}}$,
 $\left( \varphi _{\underline{5}}^d (\zeta \zeta )_{\underline{5}}\right)_{\underline{1}} (\varphi  \varphi )_{\underline{1}}$,
 $\left( \varphi _{\underline{5}}^d \big( (\varphi \varphi )_{\underline{5}} (\zeta  \zeta )_{\underline{5}} \big)_{\underline{5}} \right)_{\underline{1}}$,
 $\left( \varphi _{\underline{5}}^d \big( (\varphi \varphi )_{\underline{5}} (\zeta  \zeta )_{\underline{9}} \big)_{\underline{5}} \right)_{\underline{1}}$,
} \\
\\
 \\\hline\\[-5mm]
 
 $\chi^d_{\underline{1}}$
 & 0
 & \multirow{2}{30mm}{%
 $\chi^d_{\underline{1}} (\chi  \chi )_{\underline{1}} \eta$, \\
 $\chi^d_{\underline{1}} \bar{\eta}^3$, 
 }
 & \multirow{1}{90mm}{\footnotesize
 $\chi _{\underline{1}}^d (\xi  \xi )_{\underline{1}} (\chi  \chi )_{\underline{1}} $, 
 $\chi _{\underline{1}}^d \left( (\xi  \xi )_{\underline{5}} (\chi  \chi )_{\underline{5}} \right)_{\underline{1}} $, 
 $\chi _{\underline{1}}^d \left( (\xi  \xi )_{\underline{9}} (\chi  \chi )_{\underline{9}} \right)_{\underline{1}} $ 
} \\
 \\[1mm]\hline\\[-5mm]
 
 $\chi^d_{\underline{3}}$
 & 0
 & \multirow{2}{30mm}{%
 $\left( \chi^d_{\underline{3}} \left( \xi (\chi  \chi )_{\underline{5}} \right)_{\underline{3}} \right)_{\underline{1}}$,
 $\left( \chi^d_{\underline{3}} \left( \xi (\chi  \chi )_{\underline{9}} \right)_{\underline{3}} \right)_{\underline{1}}$}
 & \multirow{2}{90mm}{\footnotesize
 $\left(\chi _{\underline{3}}^d (\xi  \chi )_{\underline{3}}\right)_{\underline{1}} \bar{\eta} ^2$,
 $(\chi _{\underline{3}}^d \varphi )_{\underline{1}} (\chi  \chi )_{\underline{1}} \bar{\eta}$, 
 $\left( \chi _{\underline{3}}^d \big( \varphi (\chi  \chi )_{\underline{5}} \big)_{\underline{3}} \right)_{\underline{1}} \bar{\eta}$, 
 $\left( \chi _{\underline{3}}^d \big( (\chi  \chi )_{\underline{5}} (\zeta \zeta )_{\underline{5}} \big)_{\underline{3}} \right)_{\underline{1}}$, 
 } \\
 \\
 \\[3mm]\hline\\[-5mm]
 
 $\chi^d_{\underline{5}}$
 & $\left(\chi _{\underline{5}}^d (\xi  \chi )_{\underline{5}}\right)_{\underline{1}}$
 & \multirow{2}{35mm}{\footnotesize
 $\left(\chi _{\underline{5}}^d (\varphi  \chi )_{\underline{5}}\right)_{\underline{1}} \bar{\eta}$, 
 $(\chi _{\underline{5}}^d \chi )_{\underline{1}} (\zeta  \zeta )_{\underline{1}}$,
 $\left(\chi _{\underline{5}}^d \big(\chi (\zeta  \zeta )_{\underline{5}} \big)_{\underline{5}} \right)_{\underline{1}}$,
 $\left(\chi _{\underline{5}}^d \big(\chi (\zeta  \zeta )_{\underline{9}} \big)_{\underline{5}} \right)_{\underline{1}}$}
 & \multirow{1}{90mm}{\footnotesize
 $\left(\chi _{\underline{5}}^d (\xi  \chi )_{\underline{5}}\right)_{\underline{1}} \eta \bar{\eta }$,
 $\left(\chi _{\underline{5}}^d \big( \varphi (\chi \chi )_{\underline{5}} \big)_{\underline{5}} \right)_{\underline{1}} \eta$, 
 $\left(\chi _{\underline{5}}^d (\zeta \zeta )_{\underline{5}}\right)_{\underline{1}} \bar{\eta} ^2$ 
} 
 \\
 \\
 \\[11mm]\hline\\[-5mm]
 
   
 $\zeta^d_{\underline{1}}$
 & 0
 & $\zeta _{\underline{1}}^d \left(\zeta (\zeta  \zeta )_{\underline{5}}\right)_{\underline{1}}$
 & \multirow{1}{90mm}{\footnotesize
 $\zeta _{\underline{1}}^d \left(\zeta (\varphi  \chi )_{\underline{5}}\right)_{\underline{1}} \eta$
} \\[1mm]
\hline\\[-5mm]

  
 $\zeta^d_{\underline{3}}$
 & $\left(\zeta _{\underline{3}}^d (\zeta  \xi )_{\underline{3}} \right)_{\underline{1}}$
 & \footnotesize $\left(\zeta _{\underline{3}}^d (\zeta  \varphi )_{\underline{3}} \right)_{\underline{1}} \bar{\eta}$
 & \multirow{3}{90mm}{\footnotesize
 $\left( \zeta _{\underline{3}}^d (\zeta  \xi )_{\underline{3}} \right)_{\underline{1}} \eta \bar{\eta}$,
 $( \zeta _{\underline{3}}^d \varphi)_{\underline{1}} (\chi  \zeta )_{\underline{1}} \eta$,
 $\left( \zeta _{\underline{3}}^d \big( \varphi (\chi  \zeta )_{\underline{3}} \big)_{\underline{3}} \right)_{\underline{1}} \eta$,
 $\left( \zeta _{\underline{3}}^d \big( \varphi (\chi  \zeta )_{\underline{5}} \big)_{\underline{3}} \right)_{\underline{1}} \eta$,
} \\
 \\[4mm]\hline\\[-4mm]
   
 $\tilde{\zeta}^d_{\underline{1}}$
 & $\tilde{\zeta} _{\underline{1}}^d (\zeta  \zeta )_{\underline{1}}$
 & 0
 & $\tilde{\zeta} _{\underline{1}}^d (\zeta  \zeta )_{\underline{1}} \eta \bar{\eta}$,
 $\tilde{\zeta} _{\underline{1}}^d \left( (\xi  \xi )_{\underline{5}} (\varphi \chi)_{\underline{5}} \right)_{\underline{1}}$ \\
   \\[-4mm]\hline\hline

\end{tabular}
  \caption{\label{tab:flavon_superpotential} All terms up to quintic couplings in the flavon superpotential allowed by the flavour symmetry $SO(3) \times U(1)$. $\mu_\eta$ and $A_\xi$ are free parameters with one mass unit to balance the dimension in the superpotential}
\end{table}


We first list terms in the flavon superpotential which cannot be avoided by the flavour symmetry $SO(3) \times U(1)$. The full flavon superpotential should be given by
\begin{eqnarray}
w_f = w_f^{d \leqslant 3} + w_f^{d = 4} + w_f^{d = 5} + \cdots\,.
\end{eqnarray}
$w_f^{d \leqslant 3}$ represents renormalisable terms in the superpotential, and $w_f^{d = 4}$ and $w_f^{d = 5}$ are non-renormalisable quartic and quintic couplings, respectively. 
Up to quintic couplings, all terms are listed in Table~\ref{tab:flavon_superpotential}, classified by the driving fields. As mentioned above, we follow the general arrangement in most supersymmetric models that driving fields always linearly couple to flavon fields. Compared with  Eq.~\eqref{eq:flavon_superpotential}, a lot of new terms appear here. We will discuss how they modify the VEVs of $\xi$, $\varphi$, $\chi$ and $\zeta$ in detail. 

First for $\xi$, couplings involving the driving field $\xi^d_{\underline{5}}$ include not just the renormalisable term $\left( \xi _{\underline{5}}^d (\xi  \xi)_{\underline{5}} \right)_{\underline{1}}$, but also the quartic term $\left( \xi _{\underline{5}}^d (\xi \varphi)_{\underline{5}} \right)_{\underline{1}} \bar{\eta}$ and the quintic term $\left( \xi _{\underline{5}}^d \chi \right)_{\underline{1}} \eta ^3$, $\left( \xi _{\underline{5}}^d (\xi  \xi)_{\underline{5}} \right)_{\underline{1}} \eta  \bar{\eta } $, etc. 
The minimisation $\partial w_f / \partial \xi^d_{\underline{5}}= 0$ does not lead to $(\xi \xi)_{\underline{5}} = 0$, but 
\begin{eqnarray}
(\xi \xi)_{\underline{5}} = \frac{1}{\Lambda} \left(\xi \varphi \right)_{\underline{5}} \bar{\eta} + 
\frac{1}{\Lambda} \chi (\varphi  \varphi)_{\underline{1}} + \frac{1}{\Lambda^2} \chi \eta ^3 +
\frac{1}{\Lambda^2} (\xi  \xi)_{\underline{5}} \eta \bar{\eta} + \cdots \,,
\label{eq:w_xi_quintet_NLO}
\end{eqnarray}
where the dots represent contribution of all rest terms involving $\xi^d_{\underline{5}}$ in Table~\ref{tab:flavon_superpotential}. Dimensionless free parameters are omitted here and in the following. Couplings involving the driving field $\xi^d_{\underline{1}}$ is modified into
\begin{eqnarray}
(\xi \xi)_{\underline{1}} - A_\xi \eta = \frac{1}{\Lambda} \eta^2 \bar{\eta} + \frac{1}{\Lambda^2} (\xi \xi)_{\underline{1}} \eta \bar{\eta} + \cdots \,,
\label{eq:w_xi_singlet_NLO}
\end{eqnarray}
where the dots represent contributions of the rest terms in Table~\ref{tab:flavon_superpotential}. 
We denote the shifted VEV as 
\begin{eqnarray}
\xi^{A_4} + \delta_\xi\,,
\end{eqnarray}
where $\xi^{A_4}$ is the $A_4$-invariant part with each components given in Eq.~\eqref{eq:flavon_VEV} and $\delta_\xi$ represents $A_4$-breaking corrections.  Eq.~\eqref{eq:w_xi_singlet_NLO} only gives an all overall small correction to $v_\xi$ without breaking the $A_4$ symmetry. Eq.~\eqref{eq:w_xi_quintet_NLO} is approximately simplified to 
\begin{eqnarray}
2 (\delta_\xi \xi^{A_4})_{\underline{5}} \approx 
\frac{1}{\Lambda} \left(\xi^{A_4} \varphi^{Z_3} \right)_{\underline{5}} \bar{\eta} + 
\frac{1}{\Lambda} \chi^{Z_2} (\varphi^{Z_3}  \varphi^{Z_3})_{\underline{1}} + 
\frac{1}{\Lambda^2} \chi^{Z_2} \eta ^3 +
\frac{1}{\Lambda^2} (\xi^{A_4}  \xi^{A_4})_{\underline{5}} \eta \bar{\eta} + \cdots \,,
\label{eq:w_xi_quintet_NLO2}
\end{eqnarray}
where $\left((\xi^{A_4}+\delta_\xi) (\xi^{A_4}+\delta_\xi)\right)_{\underline{5}} \approx 2 (\delta_\xi \xi^{A_4})_{\underline{5}}$ is used on the left hand side and $\xi$, $\chi$ and $\varphi$ are replaced by the $A_4$-, $Z_2$- and $Z_3$-invariant VEVs $\xi^{A_4}$, $\chi^{Z_2}$ and $\varphi^{Z_3}$ on the right hand side of Eq.~\eqref{eq:flavon_VEV}, respectively. 
In our paper, since we only care about the order of magnitude of corrections, we neglect CG coefficients in the products and do a naive estimation of the order of magnitude. Then we obtain
\begin{eqnarray}
\frac{\delta_\xi}{v_\xi} \lesssim \max \{\frac{v_\varphi v_{\bar{\eta}}}{\Lambda v_\xi}, \frac{v_\chi v_\varphi^2 }{\Lambda v_\xi^2} ,  \frac{v_\chi v_\eta ^3 }{\Lambda^2 v_\xi^2}, 0, \cdots \} 
= \frac{v_\varphi v_{\bar{\eta}}}{\Lambda v_\xi}\,,
\label{eq:w_xi_quintet_NLO3}
\end{eqnarray}
where the fourth term in the curly bracket has a vanishing contribution since $(\xi^{A_4} \xi^{A_4})_{\underline{5}} = 0$. The relation in Eq.~\eqref{eq:VEV_maglitude} has been used.  
In the above estimation, we include all corrections from Table~\ref{tab:flavon_superpotential} and pick the largest one $v_\xi v_\varphi/(\Lambda v_\xi)$. Since $v_\chi, v_\varphi \ll v_\xi \ll \Lambda$, this correction is very small and can be safely ignored. The exact correction may be different from the estimation but must be smaller than it. 

Similarly, we can estimate corrections to the VEVs of $\varphi$, $\chi$ and $\zeta$. We denote the shifted VEVs of $\varphi$, $\chi$ and $\zeta$ as 
\begin{eqnarray}
& \varphi^{Z_3} + \delta_\varphi \,, \nonumber\\
& \chi^{Z_2} + \delta_\chi \,, \nonumber\\
& \zeta^{\mathbf{1}'} + \delta_\zeta \,,
\end{eqnarray}
respectively, where $\varphi^{Z_3}$, $\chi^{Z_2}$ and $\zeta^{\mathbf{1}'}$ represent leading-order value in Eq.~\eqref{eq:flavon_VEV} and $\delta_\varphi$, $\delta_\chi$ and $\delta_\zeta$ are subleading order corrections. 
Once subleading high dimensional operators are included, the minimisation of the superpotential gives rise to
\begin{eqnarray}
&&\frac{f_2}{\Lambda} \left(\delta_\xi (\varphi^{Z_3} \varphi^{Z_3})_{\underline{5}} \right)_{\underline{5}} + \frac{2f_2}{\Lambda} \left(\xi^{A_4} (\delta_\varphi \varphi^{Z_3})_{\underline{5}} \right)_{\underline{5}} 
\approx 
\frac{1}{\Lambda^2} \big( (\xi^{A_4} \xi^{A_4} )_{\underline{5}} \varphi^{Z_3} \big)_{\underline{5}} \eta \nonumber\\
&& \hspace{3cm} + \frac{1}{\Lambda^2} (\varphi^{Z_3} \varphi^{Z_3} )_{\underline{5}} (\zeta^{\mathbf{1}'}  \zeta^{\mathbf{1}'} )_{\underline{1}} + \cdots \,; \nonumber\\
&&\frac{g_2}{\Lambda}(\delta_\xi (\chi^{Z_2} \chi^{Z_2})_{\underline{5}})_{\underline{3}} + \frac{2 g_2}{\Lambda}(\xi^{A_4} (\delta_\chi \chi^{Z_2})_{\underline{5}})_{\underline{3}} + 
\frac{g_3}{\Lambda}(\delta_\xi (\chi^{Z_2} \chi^{Z_2})_{\underline{9}})_{\underline{3}} + \frac{2g_3}{\Lambda}(\xi^{A_4} (\delta_\chi \chi^{Z_2})_{\underline{9}})_{\underline{3}}  \nonumber\\ 
&& \hspace{3cm}
\approx \frac{1}{\Lambda^2} (\xi^{A_4} \chi^{Z_2})_{\underline{3}} \bar{\eta}^2 +
\frac{1}{\Lambda^2} \varphi (\chi^{Z_2}  \chi^{Z_2} )_{\underline{1}} \bar{\eta} + \cdots \,, \nonumber\\
&&(\delta_\xi \chi^{Z_2})_{\underline{5}} + (\xi^{A_4} \delta_\chi)_{\underline{5}} \approx 
\frac{1}{\Lambda} (\varphi^{Z_3}  \chi^{Z_2} )_{\underline{5}} \bar{\eta} +
\frac{1}{\Lambda} \chi^{Z_2} (\zeta^{\mathbf{1}'}  \zeta^{\mathbf{1}'} )_{\underline{1}} \cdots \,; \nonumber\\
&&h_2 (\zeta^{\mathbf{1}'}  \delta_\xi )_{\underline{3}} + h_2 (\delta_\zeta  \xi^{A_4} )_{\underline{3}} \approx
\frac{1}{\Lambda} (\zeta^{\mathbf{1}'}  \varphi^{Z_3} )_{\underline{3}} \bar{\eta} + \cdots \,, \nonumber\\
&&2 h_3 (\zeta^{\mathbf{1}'}  \delta_\zeta )_{\underline{1}} \approx
\frac{1}{\Lambda^2} (\zeta^{\mathbf{1}'} \zeta^{\mathbf{1}'} )_{\underline{1}} \eta \bar{\eta} + \cdots \,.
\label{eq:w_varphi_NLO2}
\end{eqnarray}
A naive estimation gives the upper bounds of corrections
\begin{eqnarray}
\frac{\delta_\varphi}{v_\varphi} 
&\lesssim& \max \{ \frac{\delta_\xi}{v_\xi}, 0, \frac{v_{\zeta}^2}{\Lambda v_{\xi}}, \cdots  \} =
\frac{\delta_\xi}{v_\xi} \lesssim \frac{v_\varphi v_{\bar{\eta}}}{\Lambda v_\xi} \,, \nonumber\\
\frac{\delta_\chi}{v_\chi} 
&\lesssim& \max \{ \frac{\delta_{\xi}}{v_\xi}, \frac{v_{\bar{\eta}}^2}{\Lambda v_\chi}, \frac{v_\varphi v_{\bar{\eta}}}{\Lambda v_\xi}, \cdots \} 
= \frac{v_{\bar{\eta}}^2}{\Lambda v_\chi} \,, \nonumber\\
\frac{\delta_\zeta}{v_\zeta} 
&\lesssim& \max \{ \frac{\delta_{\xi}}{v_\xi}, \frac{v_\varphi v_{\bar{\eta}}}{\Lambda v_\xi}, 0, \cdots \} = \frac{v_\varphi v_{\bar{\eta}}}{\Lambda v_\xi} \,.
\label{eq:w_varphi_quintet_NLO3}
\end{eqnarray}
Again, $(\xi^{A_4} \xi^{A_4} )_{\underline{5}}=0$, as well as $(\zeta^{\mathbf{1}'} \zeta^{\mathbf{1}'} )_{\underline{1}}=0$, and the relation in Eq.~\eqref{eq:VEV_maglitude} are used in the above. Upper bounds of relevant corrections to the $Z_3$-invariant VEV $\delta_\varphi / v_\varphi$ and the $\zeta$ VEV $\delta_\zeta / v_\zeta$ are as small as $\delta_\xi / v_\xi$. The upper bound of the correction to the $\chi$ VEV is larger, $\delta_\chi / v_\chi \lesssim v_{\bar{\eta}}^2/ (\Lambda v_\chi) \sim \sqrt{v_\eta v_{\bar{\eta}}}/\Lambda$. However, we calculate this correction in detail in Appendix~\ref{app:Z2_correction} and find that the true correction
\begin{eqnarray}
\frac{\delta_\chi}{v_\chi} \sim \frac{v_\varphi v_{\bar{\eta}}}{\Lambda v_\xi} \,,
\end{eqnarray}
which is also very small.


We numerically give an example of the size of these corrections. By setting 
\begin{eqnarray}
A_\xi = 0.3 \Lambda \,, \ \
v_{\eta} = 0.1 \Lambda \,, \ \
v_{\bar{\eta}} = 0.03 \Lambda\,, 
\end{eqnarray}
we obtain 
\begin{eqnarray}
v_\xi \sim 0.1 \Lambda\,, \ \
v_\varphi \sim 0.01 \Lambda\,, \ \
v_\chi \sim 0.03\Lambda\,, \ \
v_\zeta \sim 0.001 \Lambda \,,
\label{eq:numerical_scale}
\end{eqnarray}
and
\begin{eqnarray}
\frac{\delta_\xi}{v_\xi}, \ \frac{\delta_\varphi}{v_\varphi}, \ \frac{\delta_\zeta}{v_\zeta} \lesssim 0.005 \,, \ \
\frac{\delta_\chi}{v_\chi} \sim 0.005 \,.
\end{eqnarray}
All corrections are less than 1\%. Therefore, VEVs of $\xi$, $\chi$, $\varphi$ and $\zeta$ are stable under subleading corrections.

\subsection{Subleading corrections to flavour mixing (are important) \label{sec:subleadingflav}}

At leading order, the flavour mixing appears as the tri-bimaximal pattern. Deviation arises after subleading corrections are considered. There are two origins of subleading corrections: 
subleading higher dimensional operators in superpotential terms for lepton mass generation $w_\ell$ and
higher dimensional operators in the flavon superpotential $w_f$. The second type shift the flavon VEVs and further modify the mixing. As discussed in the last subsection, these corrections in this model are less than 1\%, safely negligible. In the following, we will only discuss corrections from the first origin. 

Subleading terms contributing to $\ell e^c H_d$ up to $d\leqslant 7$ and those to $\ell R_\mu  H_d$ or $\ell R_\tau  H_d$ up to $d\leqslant 6$ include
\begin{eqnarray} 
w_{e^c} &\supset & 
\frac{1}{\Lambda^4} (\varphi \ell )_{\underline{1}} e^c \eta^3 H_d + 
\frac{1}{\Lambda^4} \left( \big(\xi (\varphi \varphi)_{\underline{5}} \big)_{\underline{3}} \ell \right)_{\underline{1}} e^c \eta H_d \,, \nonumber\\
w_{R_\mu} &\supset & 
\frac{1}{\Lambda^2} \big( (\ell R_\mu )_{\underline{5}} (\zeta \zeta )_{\underline{5}} \big)_{\underline{1}} H_d + 
\frac{1}{\Lambda^3} \left\{ 
\big( (\ell R_\mu )_{\underline{7}} \xi \big)_{\underline{1}} \eta \bar{\eta} H_d + 
\big( (\ell R_\mu )_{\underline{3}} ( \varphi \chi)_{\underline{3}} \big)_{\underline{1}} \eta H_d \right. \nonumber\\
&& + \left.
\big( (\ell R_\mu )_{\underline{5}} ( \varphi \chi)_{\underline{5}} \big)_{\underline{1}} \eta H_d +
\big( (\ell R_\mu )_{\underline{7}} ( \varphi \chi)_{\underline{7}} \big)_{\underline{1}} \eta H_d 
\right\} \,, \nonumber\\
w_{R_\tau} &\supset & 
\frac{1}{\Lambda^2}
\big( (\ell R_\tau )_{\underline{7}} \xi \big)_{\underline{1}} \eta H_d +
\frac{1}{\Lambda^3} \left\{
\big( (\ell R_\tau )_{\underline{3}} \varphi \big)_{\underline{1}} \eta \bar{\eta} H_d +
\big( (\ell R_\tau )_{\underline{5}} (\zeta \zeta)_{\underline{5}} \big)_{\underline{1}} \eta H_d  \right. \nonumber\\
&&+
\left( (\ell R_\tau )_{\underline{3}} \big( \xi (\xi \xi)_{\underline{5}} \big)_{\underline{3}} \right)_{\underline{1}} H_d + 
\left( (\ell R_\tau )_{\underline{3}} \big( \xi (\xi \xi)_{\underline{9}} \big)_{\underline{3}} \right)_{\underline{1}} H_d +
\left( (\ell R_\tau )_{\underline{7}} \xi \right)_{\underline{1}} (\xi \xi)_{\underline{1}} H_d \nonumber\\
&&+\left.
\left( (\ell R_\tau )_{\underline{7}} \big( \xi (\xi \xi)_{\underline{5}} \big)_{\underline{7}} \right)_{\underline{1}} H_d +
\left( (\ell R_\tau )_{\underline{7}} \big( \xi (\xi \xi)_{\underline{9}} \big)_{\underline{7}} \right)_{\underline{1}} H_d \right\}
 \,.
\end{eqnarray}
For terms involving only some of $\varphi$, $\xi$, $\eta$ and $\bar{\eta}$, no $Z_3$-breaking effects are included. The $Z_3$ symmetry always guarantees that the corrected effective Yukawa couplings take the forms $(1,1,1)^T$, $(1,\omega, \omega^2)^T$ and $(1,\omega^2,\omega)^T$, as in Eqs.~\eqref{eq:e_yukawa}, \eqref{eq:mu_yukawa} and \eqref{eq:tau_yukawa}, respectively. Terms breaking the $Z_3$ symmetry are those involving $\zeta$ or $\chi$. There are five terms left, 
$\big( (\ell R_\mu )_{\underline{5}} (\zeta \zeta)_{\underline{5}} \big)_{\underline{1}} H_d$, 
$\big( (\ell R_\tau )_{\underline{5}} (\zeta \zeta)_{\underline{5}} \big)_{\underline{1}} \eta H_d$, 
$\big( (\ell R_\mu )_{\underline{3}} ( \varphi \chi)_{\underline{3}} \big)_{\underline{1}} \eta H_d$,
$\big( (\ell R_\mu )_{\underline{5}} ( \varphi \chi)_{\underline{5}} \big)_{\underline{1}} \eta H_d$, and 
$\big( (\ell R_\mu )_{\underline{7}} ( \varphi \chi)_{\underline{7}} \big)_{\underline{1}} \eta H_d$. The first two terms only contribute to coupling between $\ell$ and $R_{\mu\mathbf{3}}$ or $R_{\tau\mathbf{3}}$. The rest three terms contributing to couplings between $\ell$ and $\mu^c$. Their contributions to the charged lepton mass matrix are characterised by adding a new matrix
\begin{eqnarray}
\delta M_l = \frac{ v_\eta v_\chi v_\varphi}{ \Lambda ^3}
\left(
\begin{array}{ccc}
 0 & 0 & 0 \\
 0 & c \omega +d \omega ^2 & 0 \\
 0 & c \omega ^2+d \omega  & 0 \\
\end{array}
\right) \frac{v_d}{\sqrt{2}}
\end{eqnarray}
to $M_l$. 
Acting $U_\omega^T$ on the left hand side of $\delta M_l$ leaves
\begin{eqnarray}
U_\omega^T \delta M_l = \frac{ v_\eta v_\chi v_\varphi}{\sqrt{3} \Lambda ^3}
\left(
\begin{array}{ccc}
 0 & - c - d & 0 \\
 0 & 2 c - d & 0 \\
 0 & 2 c - d & 0 \\
\end{array}
\right) \frac{v_d}{\sqrt{2}} \,,
\end{eqnarray}
where $c$ and $d$ are real dimensionless parameters. The unitary matrix to diagonalise $M_l$ is modified to $U_l \simeq U_{\rm \omega} U_{e\mu}$, where $U_{e\mu}$ is a complex rotation matrix on the $e \mu$ plane, 
\begin{eqnarray} 
U_{e\mu}=  
\begin{pmatrix}
 \cos \theta_{e\mu} & \sin \theta_{e\mu} e^{- i \phi_{e\mu}} & 0 \\
 -\sin \theta_{e\mu} e^{i \phi_{e\mu}} & \cos \theta_{e\mu} & 0 \\
 0 & 0 & 1 \\
\end{pmatrix}
\end{eqnarray}
with 
\begin{eqnarray}
\sin \theta_{e\mu} &=& \frac{(c+d) v_\eta v_\chi}{y_\mu v_{\bar{\eta}} \Lambda} \,,\nonumber\\
\phi_{e\mu} &=& \arg \big(- (c+d) v_\eta v_\chi}{y_\mu v_{\bar{\eta}} \Lambda \big) \,.
\end{eqnarray}
Here, we have ignored the $(3,2)$ entry of $U_\omega^T \delta M_l$ since it is too small compared with the $\tau$ mass $m_\tau \sim v_\varphi v_d / \Lambda$. 

In the neutrino sector, terms for neutrino masses up to $d\leqslant 5$ have only trivial corrections, 
$w_N \supset  \frac{1}{\Lambda^2} \set{
(\ell N)_{\underline{1}} \eta \bar{\eta} H_u + 
 \big( \chi (N N)_{\underline{5}} \big)_{\underline{1}} \eta \bar{\eta} } $. 
Therefore, the unitary matrix $U_\nu$ to diagonal $M_\nu$ keeps the same as that in the leading order.

Including the subleading correction, the PMNS matrix is modified into $U_{\rm PMNS} = U_{e\mu}^\dag U_{\rm TBM}$. multiplying $U_{e\mu}$ on the left hand side does not change the third row of the PMNS matrix. Three mixing angles are given by \cite{King:2005bj}
\begin{eqnarray}
\sin \theta_{13} &=& \frac{\sin\theta_{e\mu}}{\sqrt{2}} \,, \nonumber\\
\sin \theta_{12} &=& \sqrt{\frac{2-2\sin2\theta_{e\mu} \cos\phi_{e\mu}}{3 (2 - \sin^2\theta_{e\mu})}} \,, \nonumber\\
\sin \theta_{23} &=& \frac{\cos\theta_{e\mu}}{\sqrt{2 - \sin^2\theta_{e\mu}}} \,.
\end{eqnarray}
In this model, $\theta_{23}$ in the first octant is predicted. The reactor angle $\theta_{13} \sim v_\eta v_\chi / (v_{\bar{\eta}} \Lambda)$. For the numerical value in Eq.~\eqref{eq:numerical_scale}, we have $v_\eta v_\chi / (v_{\bar{\eta}} \Lambda) \sim 0.05$. In order to generate sizeable value of $\theta_{13}$, a relatively large value of the ratio $(c+d)/y_\mu$ is required. This is not hard to be achieved. 
The Dirac-type CP-violating phase is predicted to be
\begin{eqnarray}
\delta = \arg \big( (3 \cos 2 \theta_{e\mu}+\cos 4 \theta_{e\mu}) \cos \phi_{e\mu} - i (\cos 2 \theta_{e\mu}+3) \sin \phi_{e\mu}+\sin 2 \theta_{e\mu} \big) \,.
\end{eqnarray}
The unknown phase $\phi_{e\mu}$ can be eliminated to yield sum rules which have been widely studied
\cite{King:2005bj,Antusch:2005kw}.
In the limit $\phi_{e\mu} \to \pi/2$, an almost maximal CP-violating phase $\delta \sim 3\pi/2$ is predicted. 

\subsection{Phenomenological implications of gauged $SO(3)$ \label{sec:gauge_interactions}}

We label the gauge field of $SO(3)$ and $U(1)$ as $F^{\prime \, 1,2,3}$ and $B'$, respectively. Their interactions with flavons or fermions are simply obtained with the replacement 
\begin{eqnarray} 
\partial_\mu \to D_\mu = \partial_\mu + {\rm g}'_3 \sum_{a=1,2,3} F^{\prime \, a}_\mu \tau^a + Q {\rm g}'_1 B'_\mu \,. 
\end{eqnarray}
in the kinetic terms of the relevant fields. Here, ${\rm g}'_3$ and ${\rm g}'_1$ are gauge couplings of $SO(3)$ and $U(1)$, respectively, and the $U(1)$ charge $Q$ for each field is listed in Table~\ref{tab:fields}. 

Specifically, the kinetic term for $\xi$ in Eq.~\eqref{eq:kinetic_xi} is replaced by $(D_\mu \xi^* D^\mu \xi)_{\underline{1}}$ with 
\begin{eqnarray}
(D_\mu \xi)_{ijk} = (\partial_\mu \xi)_{ijk} + {\rm g}'_3 \sum_{a=1,2,3} F^{\prime \, a}_\mu \left[(\tau^a)_{il} \xi_{ljk} + (\tau^a)_{jl} \xi_{ilk} + (\tau^a)_{kl} \xi_{ijl}\right] + Q {\rm g}'_1 B'_\mu \xi_{ijk} \,. 
\label{eq:kinetic_xi_gauge}
\end{eqnarray}
where $Q=+1$ for $\xi$ has been used. $F^{\prime \ 1,2,3}_\mu$ gain masses once $\xi$ get the $A_4$-invariant VEV. We obtain that $M^2_{F^{\prime \,1}} = M^2_{F^{\prime \,2}} = M^2_{F^{\prime \,3}} = (2 {\rm g}'_3 v_\xi)^2$. The degenerate mass spectrum is also consistent with the $A_4$ symmetry \footnote{One may use the generators $s$ and $t$ to perform a $A_4$ transformation, an $A_4$-invariant mass term for $F'$ is obtained only if all masses are degenerated. }. Later after the rest flavons $\zeta$, $\varphi$ and $\chi$ gain VEVs, mass splitting are generated among $F^{\prime 1,2,3}$. Since VEVs of $\zeta$, $\varphi$ and $\chi$ are much smaller than that of $\xi$, the mass splittings are very small, and masses of $F^{\prime \,1,2,3}$ are still nearly degenerate. 

$B'$ obtains a mass from VEVs of both $\xi$ and $\eta$, $\bar{\eta}$, $M^2_{B^{\prime}} = {\rm g}^{\prime \, 2}_1 (v_\xi^2 + v_\eta^2 + v_{\bar{\eta}}^2)$. After $A_4$ breaking, VEVs of $\zeta$, $\varphi$ and $\chi$ contribute small corrections to the $B'$ mass. Interactions between leptons and $B'$ are flavour-dependent, with charges for $\ell$, $e^c$, $\mu^c$ and $\tau^c$ given by $-\frac{2}{3}$, $-\frac{7}{3}$, $-1$ and $-\frac{1}{3}$, respectively. 

In the limit of the $A_4$ invariance, there is no mixing between $F'$ and $B'$. This can be simply explained as follows. The mixing between $F'$ and $B'$ from Eq.~\eqref{eq:kinetic_xi_gauge} and $(D_\mu \xi^* D^\mu \xi)_{\underline{1}}$, if exists, can be only generated via coupling $F'B'\xi\xi$. Since $F' \sim \underline{3}$, $B' \sim \underline{1}$, $\xi \sim \underline{7}$, the only $SO(3)$ invariant formed by these fields is $B_\mu \big( F^\mu (\xi^* \xi)_{\underline{3}} \big)_{\underline{1}}$. Here, the $\underline{3}$-plet contraction between $\xi^*$ and $\xi$ are anti-symmetric. Once $\xi$ get the VEV, where only one of the seven components has a non-zero value, $\langle \xi_{0} \rangle = v_\xi$, the anti-symmetric contraction $\langle (\xi^* \xi)_{\underline{3}} \rangle$ vanishes. Therefore, there is no mixing between $F'$ and $B'$. The mixing between $F'$ and $B'$ is generated after $A_4$ breaking, induced by terms such as $B_\mu \big( F^\mu (\xi^* \chi)_{\underline{3}} \big)_{\underline{1}}$. The resulted mixing between $F'$ and $B'$ is suppressed by the ratio $v_\chi / v_\xi$. 

These gauge bosons are supposed to be very heavy, with masses around $\sim \mathcal{O}(v_\xi)$ or $\sim \mathcal{O}(\max(v_\xi, v_\eta, v_{\bar{\eta}}))$, respectively, if gauge coefficients are of order one. However, they could be much lighter if gauge couplings are tiny. For example, if $\Lambda$ is fixed at $10^4$ TeV, $v_\xi$ and $v_\eta$ are predicted to be around $10^3$ TeV and $v_\chi$, $v_\varphi$ and $v_\zeta$ be around 100 TeV. For a gauge coupling around $10^{-3}$, TeV-scale gauge bosons are predicted. Then, interesting signatures involving gauge interactions can be tested at colliders or precision measurements of charged leptons. Another interesting point is the prediction of a heavy tau lepton with mass also around TeV scale ($v_\zeta^2 / \Lambda \sim 1$ TeV). Its interaction with $B'$ can be tested at colliders. 

\subsection{Absence of domain walls}

The domain wall problem is a well-known problem for discrete symmetry breaking. In this paper, all flavour symmetries at high scale are gauged. $A_4$, and the residual symmetries $Z_3$ and $Z_2$, are just phenomenologically effective symmetries at lower scales. The usual domain wall problem for the global symmetry breaking does not apply here. 

In our model, we actually have a two-step phase transition  $SO(3) \to A_4$ and $A_4 \to Z_3, Z_2$. We discuss more on why the topological defect of domain walls does not exit in the model. 

At the first step, $SO(3) \to A_4$, the breaking of a gauge symmetry does not introduce domain walls. As noted in section~\ref{sec:SO3}, there are degenerate vacuums which are continuously connected by $SO(3)$ basis transformation as in Eq.~\eqref{eq:SO3_transformation}. All vacuums are perturbatively equivalent.

At the second step, $A_4 \to Z_3, Z_2$, degenerate $Z_3$-invariant or $Z_2$-invariant vacuums exit, as shown in Eqs.~\eqref{eq:phi_VEV} and \eqref{eq:chi_VEV}. Taking the $Z_3$-invariant vacuum as an example, different $Z_3$-invariant vacuums are randomly generated during $A_4$ breaking to $Z_3$ and domain walls separating different vacuums arise. These domain walls store energy with energy density inside the wall around $v_\varphi^4$ or $v_\chi^4$. Without considering gauge interactions, there are not enough energy inputted to force one vacuum jumping across the wall into another. Therefore, domain walls survive. Once gauge interactions are included, domain walls should decay to light particles mediated by gauge bosons. For the case of small gauge couplings, the gauge bosons may be light enough, i.e., $M_{F'} \lesssim v_\varphi$, and domain walls may directly decay into gauge bosons. 

\section{Conclusion \label{conclusion}}

In this paper we have discussed the breaking of $SO(3)$ down to finite family symmetries such as $A_4$, $S_4$ and $A_5$
using {\bf supersymmetric} potentials for the first time. We have analysed in detail the case of 
supersymmetric $A_4$ 
and its finite subgroups $Z_3$ and $Z_2$. We have proposed a supersymmetric $A_4$ model
of leptons along these lines, originating from $SO(3)\times U(1)$, which leads to a phenomenologically acceptable pattern of lepton mixing and masses once subleading corrections are taken into account. We have also discussed the phenomenological 
consequences of having a gauged $SO(3)$, leading to massive gauge bosons, and have shown that all domain wall problems are resolved in this model.

The main achievement of the paper is to show for the first time that {\bf supersymmetric} $SO(3)$ flavour symmetry can be the origin of finite non-Abelian family symmetry models. By focussing in detail on a supersymmetric $A_4$ model,
we have demonstrated that such a strategy can lead to a viable lepton model which can explain all oscillation data
with SUSY being preserved in the low energy spectrum (below the flavour symmetry
breaking scales). Moreover, we have shown that, if the $SO(3)$ is gauged, there may be interesting phenomenological
implications due to the massive gauge bosons. 

About a half of the paper is devoted to the study of the realistic supersymmetric $A_4$ model of leptons,
arising from $SO(3)\times U(1)$. This study is important in order to verify that it is really possible to 
construct a fully working model along these lines.
The main achievements of the specific model may be summarised as follows: 

\begin{itemize}

\item We have achieved the breaking of $SO(3) \to A_4$ in SUSY, using high irreps of $SO(3)$
and flat directions. In this paper, we have chosen a $\underline{7}$-plet, i.e., a rank-3 tensor in 3d space, to achieve the breaking. We have shown that it is possible to break $SO(3)$ to $S_4$ or $A_5$ by using different higher irreps.

\item We have shown that it is possible to also achieve, at the level of $SO(3)$, the subsequent breaking of 
$A_4$ at a lower scale (below the $SO(3)$ breaking scale) to the residual symmetries $Z_3$ and $Z_2$. 
Such $Z_3$ and $Z_2$ symmetries are preserved in charged lepton sector and neutrino sector, respectively, 
after the $A_4$ breaking, in accordance with the semi-direct model building strategy.

\item Starting from a supersymmetric flavour group $SO(3)\times U(1)$, we have shown how 
$SO(3)$ is broken first to $A_4$, and then to $Z_3$ and $Z_2$. The $A_4$, $Z_3$ and $Z_2$ symmetries are respectively achieved by the flavons $\xi$, $\varphi$ and $\chi$ after they gain the $A_4$-, $Z_3$- and $Z_2$-invariant VEVs, respectively. 
We have found that tri-bimaximal mixing (with zero reactor angle) is realised at leading order. One technical point is that the singlet irreps $\mathbf{1}'$ and $\mathbf{1}''$ of $A_4$ always accompany each other after $SO(3)$ breaking. To avoid any fine tuning of parameters related to $\mu$ and $\tau$ masses, we have introduced an additional flavon $\zeta$ to split the $\mathbf{1}'$ and $\mathbf{1}''$.

\item We have considered the influence of the higher dimensional operator corrections to the model. 
We have shown that the $A_4$-, $Z_3$- and $Z_2$-invariant VEVs are stable even after subleading corrections are included. 
However, we have seen that the charged lepton mass matrix is modified by higher dimensional operators, due to the coupling with $\chi$, which gains the $Z_2$-invariant VEV. This welcome correction leads to additional mixing between $e$ and $\mu$, giving rise to a non-zero $\theta_{13}$ and the CP-violating phase $\delta$. 

\item If the $SO(3)\times U(1)$ is gauged, the model predicts three gauge bosons $F^{\prime 1,2,3}$ with the nearly degenerate masses after $SO(3)$ breaking to $A_4$. Another gauge boson $B'$ gain a mass after $U(1)$ is broken. These gauge bosons with their flavour-dependent interactions with leptons will lead to phenomenological signatures worthy of further study. 

\item We emphasise that the flavour symmetry at high scale is the continuous gauge symmetry $SO(3) \times U(1)$, 
with no {\it ad hoc} discrete symmetries introduced, and $A_4$ being just an effective flavour symmetry below the $SO(3)$ breaking 
scale. We have shown that the usual domain wall problems encountered in $A_4$ models are resolved here. 

\end{itemize}


\subsection*{Acknowledgements}

S.\,F.\,K.\ acknowledges the STFC Consolidated Grant ST/L000296/1 and the European Union's Horizon 2020 Research and Innovation programme under Marie Sk\l{}odowska-Curie grant agreements Elusives ITN No.\ 674896 and InvisiblesPlus RISE No.\ 690575. Y.\,L.\,Z.\ is supported by European Research Council under ERC Grant NuMass (FP7-IDEAS-ERC ERC-CG 617143). Y.\,L.\,Z.\ thanks Luca Di Luzio for a useful discussion.


\appendix
\section{Clebsch-Gordan coefficients of $SO(3)$ \label{sec:CG}}

In $SO(3)$, the product of two irreducible representations (irreps) $\phi$ of dimension $2p+1$
and $\Psi$ of dimension $2q+1$ are decomposed as follows:
\begin{eqnarray}
(\underline{2p\!+\!1}) \times (\underline{2q\!+\!1}) = (\underline{2|p\!-\!q|\!+\!1}) + (\underline{2|p\!-\!q|\!+\!3}) + \cdots + (\underline{2(p\!+\!q)\!+\!1})
\end{eqnarray}
Some useful Clebsch-Gordan coefficients of these products in the 3d space are listed in the following: 
\begin{itemize}
\item For $\phi \sim \Psi \sim \underline{3}$, $\underline{3} \times \underline{3}= \underline{1} + \underline{3} + \underline{5}  $, 
\begin{eqnarray}
(\phi\Psi)_{\underline{1}} ~~~ &\sim& \phi_{a} \Psi_{a} \,, \nonumber\\
\big((\phi\Psi)_{\underline{3}}\big)_{i} &\sim& \epsilon _{i a b} \phi_{a} \Psi_{b} \,, \nonumber\\
\big((\phi\Psi)_{\underline{5}}\big)_{ij} &\sim& \phi_{i a} \Psi_{j a} -\frac{1}{3} \delta _{i j} \phi_{a} \Psi_{a}+(\text{perms of $ij$}) \,. 
\end{eqnarray}

\item For $\phi \sim \underline{3}$ and $ \Psi \sim \underline{5}$, $\underline{3} \times \underline{5}= \underline{3} + \underline{5} + \underline{7}  $, 
\begin{eqnarray}
\big((\phi\Psi)_{\underline{3}}\big)_{i} &\sim& \phi_{a} \Psi_{i a} \,, \nonumber\\
\big((\phi\Psi)_{\underline{5}}\big)_{ij} &\sim& \epsilon_{i a b}\phi_{a} \Psi_{j b} +(\text{perms of $ij$}) \,, \nonumber\\
\big((\phi\Psi)_{\underline{7}}\big)_{ijk} &\sim& \phi_{i} \Psi_{j k} - \frac{2}{5} \delta_{ij} \phi_a \Psi_{k a}  +(\text{perms of $ijk$}) \,.
\end{eqnarray}

\item For $\phi \sim \underline{3}$, $\Psi \sim \underline{7}$, $\underline{3} \times \underline{7}= \underline{5} + \underline{7} + \underline{9}  $, 
\begin{eqnarray}
\big((\phi\Psi)_{\underline{5}}\big)_{ij}  &\sim& \phi_{a} \Psi_{i j a} +(\text{perms of $ij$}) \,, \nonumber\\
\big((\phi\Psi)_{\underline{7}}\big)_{ijk} &\sim& \epsilon _{i a b} \phi_{a} \Psi_{j k b} +(\text{perms of $ijk$}) \,, \nonumber\\
\big((\phi\Psi)_{\underline{9}}\big)_{ijkl} &\sim& \phi_{i} \Psi_{j k l} -\frac{3}{7} \delta _{i j} \phi_{a} \Psi_{k l a}+(\text{perms of $ijkl$}) \,. 
\end{eqnarray}

\item For $\phi \sim \underline{3}$, $ \Psi \sim \underline{9}$, $\underline{3} \times \underline{9}= \underline{7} + \underline{9} + \underline{11}  $, 
\begin{eqnarray}
\big((\phi\Psi)_{\underline{7}}\big)_{ijk} &\sim& \phi_{a} \Psi_{i j k a} +(\text{perms of $ijk$}) \,, \nonumber\\
\big((\phi\Psi)_{\underline{9}}\big)_{ijkl} &\sim& \epsilon_{i a b}\phi_{a} \Psi_{j k l b} +(\text{perms of $ijkl$}) \,, \nonumber\\
\big((\phi\Psi)_{\underline{11}}\big)_{ijklm} &\sim& \phi_{i} \Psi_{j k l m} - \frac{4}{9} \delta_{ij} \phi_a \Psi_{k l m a}  +(\text{perms of $ijklm$}) \,.
\end{eqnarray}

\item For $\phi \sim \Psi \sim \underline{5}$, $\underline{5} \times \underline{5}= \underline{1} + \underline{3} + \underline{5} + \underline{7} + \underline{9} $, 
\begin{eqnarray}
(\phi\Psi)_{\underline{1}} ~~~ &\sim& \phi_{a b} \Psi_{a b} \,, \nonumber\\
\big((\phi\Psi)_{\underline{3}}\big)_{i} &\sim& \epsilon _{i a b} \phi_{a c} \Psi_{b c} \,, \nonumber\\
\big((\phi\Psi)_{\underline{5}}\big)_{ij} &\sim& \phi_{i a} \Psi_{j a} -\frac{1}{3} \delta _{i j} \phi_{a b} \Psi_{a b}+(\text{perms of $ij$}) \,, \nonumber\\
\big((\phi\Psi)_{\underline{7}}\big)_{ijk} &\sim& \epsilon _{i a b} \phi_{j a} \Psi_{k b} - \frac{1}{5} \epsilon _{i a b} \delta _{j k} \phi_{a c} \Psi_{b c} +(\text{perms of $ijk$}) \,, \nonumber\\
\big((\phi\Psi)_{\underline{9}}\big)_{ijkl} &\sim& \phi_{i j} \Psi_{k l} - \frac{4}{7} \delta _{i j} \phi_{k a} \Psi_{l a} + \frac{2}{35} \delta _{i j} \delta _{k l} \phi_{a b} \Psi_{a b} +(\text{perms of $ijkl$}) \,. 
\end{eqnarray}

\item For $\phi \sim \underline{5}$, $ \Psi \sim \underline{7}$, $\underline{5} \times \underline{7}= \underline{3} + \underline{5} + \underline{7} + \underline{9} + \underline{11} $, 
\begin{eqnarray}
\big((\phi\Psi)_{\underline{3}}\big)_{i} &\sim& \phi_{a b} \Psi_{i a b} \,, \nonumber\\
\big((\phi\Psi)_{\underline{5}}\big)_{ij} &\sim& \epsilon_{i a b}\phi_{a c} \Psi_{j b c} +(\text{perms of $ij$}) \,, \nonumber\\
\big((\phi\Psi)_{\underline{7}}\big)_{ijk} &\sim& \phi_{i a} \Psi_{j k a} - \frac{2}{5} \delta_{ij} \phi_{a b} \Psi_{k a b}  +(\text{perms of $ijk$}) \,, \nonumber\\
\big((\phi\Psi)_{\underline{9}}\big)_{ijkl} &\sim& \epsilon_{i a b}\phi_{j a} \Psi_{k l b} - \frac{2}{7} \epsilon_{i a b} \delta_{jk} \phi_{a c} \Psi_{l b c} +(\text{perms of $ijkl$}) \,, \nonumber\\
\big((\phi\Psi)_{\underline{11}}\big)_{ijkl} &\sim& \phi_{i j} \Psi_{k l m} - \frac{2}{3} \delta_{ij} \phi_{k a} \Psi_{l m a} + \frac{2}{21} \delta_{ij} \delta_{kl} \phi_{a b} \Psi_{m a b} +(\text{perms of $ijkl$}) \,. \nonumber\\
\end{eqnarray}

\item For $\phi \sim \Psi \sim \underline{7}$, $\underline{7} \times \underline{7} = \underline{1} + \underline{3} + \underline{5} + \underline{7} + \underline{9} + \underline{11} + \underline{13} $, 
\begin{eqnarray}
(\phi\Psi)_{\underline{1}} ~~~ &\sim& \phi_{a b c} \Psi_{a b c} \,, \nonumber\\
\big((\phi\Psi)_{\underline{3}}\big)_{i} &\sim& \epsilon _{i a b} \phi_{a c d} \Psi_{b c d} \,, \nonumber\\
\big((\phi\Psi)_{\underline{5}}\big)_{ij} &\sim& \phi_{i a b} \Psi_{j a b} -\frac{1}{3} \delta _{i j} \phi_{a b c} \Psi_{a b c}+(\text{perms of $ij$}) \,, \nonumber\\
\big((\phi\Psi)_{\underline{7}}\big)_{ijk} &\sim& \epsilon _{i a b} \phi_{j a c} \Psi_{k b c} - \frac{1}{5} \epsilon _{i a b} \delta _{j k} \phi_{a c d} \Psi_{b c d} +(\text{perms of $ijk$}) \,, \nonumber\\
\big((\phi\Psi)_{\underline{9}}\big)_{ijkl} &\sim& \phi_{i j a} \Psi_{k l a} - \frac{4}{7} \delta _{i j} \phi_{k a b} \Psi_{l a b} + \frac{2}{35} \delta _{i j} \delta _{k l} \phi_{a b c} \Psi_{a b c} +(\text{perms of $ijkl$}) \,, \nonumber\\
\big((\phi\Psi)_{\underline{11}}\big)_{ijklm} &\sim& \epsilon _{i a b} \phi_{j k a} \Psi_{l m b} - \frac{4}{9} \epsilon _{i a b} \delta _{j k} \phi_{l a c} \Psi_{m b c} + \frac{2}{63} \epsilon _{i a b} \delta _{j k} \delta _{l m} \phi_{a c d} \Psi_{b c d} \nonumber\\
&& +(\text{perms of $ijklm$}) \,, \nonumber\\
\big((\phi\Psi)_{\underline{13}}\big)_{ijklmn} &\sim& \phi_{i j k} \Psi_{l m n} - \frac{9}{11} \delta _{i j} \phi_{k l a} \Psi_{m n a} + \frac{2}{11} \delta _{i j} \delta _{k l} \phi_{m a b} \Psi_{n a b} \nonumber\\ 
&& - \frac{2}{231} \delta _{i j} \delta _{k l} \delta _{m n} \phi_{a b c} \Psi_{a b c} +(\text{perms of $ijklmn$}) \,.
\end{eqnarray}

\item For $\phi \sim \underline{7}$, $ \Psi \sim \underline{9}$, $\underline{7} \times \underline{9}= \underline{3} + \underline{5} + \underline{7} + \underline{9} + \underline{11} + \underline{13} + \underline{15} $, 
\begin{eqnarray}
\big((\phi\Psi)_{\underline{3}}\big)_{i} &\sim& \phi_{a b c} \Psi_{i a b c} \,, \nonumber\\
\big((\phi\Psi)_{\underline{5}}\big)_{ij} &\sim& \epsilon_{i a b}\phi_{a c d} \Psi_{j b c d} +(\text{perms of $ij$}) \,, \nonumber\\
\big((\phi\Psi)_{\underline{7}}\big)_{ijk} &\sim& \phi_{i a b} \Psi_{j k a b} - \frac{2}{5} \delta_{ij} \phi_{a b c} \Psi_{k a b c}  +(\text{perms of $ijk$}) \,, \nonumber\\
\big((\phi\Psi)_{\underline{9}}\big)_{ijkl} &\sim& \epsilon_{i a b}\phi_{j a c} \Psi_{k l b c} - \frac{2}{7} \epsilon_{i a b} \delta_{jk} \phi_{a c d} \Psi_{l b c d} +(\text{perms of $ijkl$}) \,, \nonumber\\
\big((\phi\Psi)_{\underline{11}}\big)_{ijkl} &\sim& \phi_{i j a} \Psi_{k l m a} - \frac{2}{3} \delta_{ij} \phi_{k a b} \Psi_{l m a b} + \frac{2}{21} \delta_{ij} \delta_{kl} \phi_{a b c} \Psi_{m a b c} \nonumber\\
&& +(\text{perms of $ijkl$}) \,.
\end{eqnarray}

\item For $\phi \sim \Psi \sim \underline{9}$, $\underline{9} \times \underline{9} = \underline{1} + \underline{3} + \underline{5} + \underline{7} + \underline{9} + \underline{11} + \underline{13} + \underline{15} + \underline{17} $, 
\begin{eqnarray}
(\phi\Psi)_{\underline{1}} ~~~ &\sim& \phi_{a b c d} \Psi_{a b c d} \,, \nonumber\\
\big((\phi\Psi)_{\underline{3}}\big)_{i} &\sim& \epsilon _{i a b} \phi_{a c d f} \Psi_{b c d f} \,, \nonumber\\
\big((\phi\Psi)_{\underline{5}}\big)_{ij} &\sim& \phi_{i a b c} \Psi_{j a b c} -\frac{1}{3} \delta _{i j} \phi_{a b c d} \Psi_{a b c d}+(\text{perms of $ij$}) \,, \nonumber\\
\big((\phi\Psi)_{\underline{7}}\big)_{ijk} &\sim& \epsilon _{i a b} \phi_{j a c d} \Psi_{k b c d} - \frac{1}{5} \epsilon _{i a b} \delta _{j k} \phi_{a c d f} \Psi_{b c d f} +(\text{perms of $ijk$}) \,, \nonumber\\
\big((\phi\Psi)_{\underline{9}}\big)_{ijkl} &\sim& \phi_{i j a b} \Psi_{k l a b} - \frac{4}{7} \delta _{i j} \phi_{k a b c} \Psi_{l a b c} + \frac{2}{35} \delta _{i j} \delta _{k l} \phi_{a b c d} \Psi_{a b c d} \nonumber\\
&& +(\text{perms of $ijkl$}) \,, \nonumber\\
\big((\phi\Psi)_{\underline{11}}\big)_{ijklm} &\sim& \epsilon _{i a b} \phi_{j k a c} \Psi_{l m b c} - \frac{4}{9} \epsilon _{i a b} \delta _{j k} \phi_{l a c d} \Psi_{m b c d} \nonumber\\
&& + \frac{2}{63} \epsilon _{i a b} \delta _{j k} \delta _{l m} \phi_{a c d f} \Psi_{b c d f} +(\text{perms of $ijklm$}) \,, \nonumber\\
\big((\phi\Psi)_{\underline{13}}\big)_{ijklmn} &\sim& \phi_{i j k a} \Psi_{l m n a} - \frac{9}{11} \delta _{i j} \phi_{k l a b} \Psi_{m n a b} + \frac{2}{11} \delta _{i j} \delta _{k l} \phi_{m a b c} \Psi_{n a b c} \nonumber\\ 
&& - \frac{2}{231} \delta _{i j} \delta _{k l} \delta _{m n} \phi_{a b c d} \Psi_{a b c d} +(\text{perms of $ijklmn$}) \,.
\end{eqnarray}

\item For $\phi \sim \Psi \sim \underline{13}$, $\underline{13} \times \underline{13} = \underline{1} + \underline{3} + \underline{5} + \underline{7} + \underline{9} + \underline{11} + \underline{13} + \underline{15} + \underline{17} + \underline{19} + \underline{21} + \underline{23} + \underline{25} $, 
\begin{eqnarray}
(\phi\Psi)_{\underline{1}} ~~~ &\sim& \phi_{a b c d f g} \Psi_{a b c d f g} \,, \nonumber\\
\big((\phi\Psi)_{\underline{5}}\big)_{ij} &\sim& \phi_{i a b c d f} \Psi_{j a b c d f} -\frac{1}{3} \delta _{i j} \phi_{a b c d f g} \Psi_{a b c d f g}+(\text{perms of $ij$}) \,, \nonumber\\
\big((\phi\Psi)_{\underline{9}}\big)_{ijkl} &\sim& \phi_{i j a b c d} \Psi_{k l a b c d} - \frac{4}{7} \delta _{i j} \phi_{k a b c d f} \Psi_{l a b c d f} + \frac{2}{35} \delta _{i j} \delta _{k l} \phi_{a b c d f g} \Psi_{a b c d f g} \nonumber\\
&& +(\text{perms of $ijkl$}) \,.
\end{eqnarray}

\end{itemize}

\section{Solutions of the superpotential minimisation \label{app:minimisation}}

\subsection{Solutions for $SO(3) \to A_4$}
Equations for the minimisation of the superpotential term $w_\xi$ in Eqs.~\eqref{eq:w_xi_singlet} and \eqref{eq:w_xi_quintet} are respectively and explicitly written out as
\begin{eqnarray}
\hspace{-5mm}-\frac{\mu_\xi^2}{c_1} + 2 \xi_{111}^2+3 \xi_{111} \xi_{133}+2 \xi_{112}^2+\xi_{112} \xi_{233}+3 \xi_{113}^2+3 \xi_{113} \xi_{333}&& \nonumber\\
+3 \xi_{123}^2+3 \xi_{133}^2+2 \xi_{233}^2+2 \xi_{333}^2 &=& 0 \,; \label{eq:w_xi_singlet_v2} \\
2 \left(\xi_{111}^2+\xi_{112}^2-\xi_{112} \xi_{233}-3 \xi_{113} \xi_{333}-2 \xi_{233}^2-2 \xi_{333}^2\right) &=& 0 \,, \nonumber\\
3 \xi_{111} \xi_{233}-3 \xi_{112} \xi_{133}-6 \xi_{123} \xi_{333}+6 \xi_{133} \xi_{233} &=& 0 \,, \nonumber\\
3 \xi_{111} (2 \xi_{113}+\xi_{333})+6 \xi_{112} \xi_{123}+9 \xi_{113} \xi_{133}+6 \xi_{123} \xi_{233}+6 \xi_{133} \xi_{333} &=& 0 \,, \nonumber\\
-6 \xi_{111} \xi_{123}+6 \xi_{112} \xi_{113}+3 \xi_{112} \xi_{333}-3 \xi_{113} \xi_{233} &=& 0 \,, \nonumber\\
2 \left(-2 \xi_{111}^2-3 \xi_{111} \xi_{133}-2 \xi_{112}^2-\xi_{112} \xi_{233}+\xi_{233}^2+\xi_{333}^2\right) &=& 0 \,. \label{eq:w_xi_quintet_v2} 
\end{eqnarray}
Five equations in Eq.~\eqref{eq:w_xi_quintet_v2} corresponds to it (11), (12), (13), (23) and (33) entries of two rank-2 tensor $(\xi \xi)_{\underline{5}} \equiv \partial w_\xi / (c_2 \partial \xi^d_{\underline{5}})$, respectively. 
By setting $\xi_{111}=\xi_{112}=\xi_{113}=\xi_{133}=\xi_{233}=\xi_{333}=0$, Eq.~\eqref{eq:w_xi_quintet_v2} is automatically satisfied. Then, Eq.~\eqref{eq:w_xi_singlet_v2} is left with 
\begin{eqnarray}
-\frac{\mu_\xi^2}{c_1} + 3 \xi_{123}^2 = 0 \,,
\end{eqnarray}
from which we obtain $\xi_{123} = \pm \sqrt{\mu_\xi^2/(3c_1)}$. Then, we arrive at the special solution in Eq.~\eqref{eq:xi_VEV}. 

\subsection{Solutions for $A_4 \to Z_3$}

Equations for the minimasation of $w_\varphi$ is given in Eq.~\eqref{eq:w_phi}. 
Taking the VEV of $\xi$ in Eq.~\eqref{eq:xi_VEV} into these equations, i.e., $\xi_{111}=\xi_{112}=\xi_{113}=\xi_{133} = \xi_{233} = \xi_{333} = 0$, part of these equations are automatically satisfied, the left vanishing ones are  simplified as
\begin{eqnarray}
\varphi_1^2+\varphi_2^2+\varphi_3^2 - \frac{\mu_\varphi ^2}{f_1} &=& 0 \,, \nonumber \\
4 \xi_{123} \left(\varphi_2^2-\varphi_3^2\right) &=& 0 \,, \nonumber \\
-4 \xi_{123} \left(\varphi_2^2-\varphi_1^2\right) &=& 0 \,.
\end{eqnarray}
It is straightforward to derive all solutions in Eq.~\eqref{eq:phi_VEV}.

\subsection{Solutions for $A_4 \to Z_2$}

Equations of minimisation of $w_\chi$ are given in Eq.~\eqref{eq:w_chi}. After $\xi$ get the $A_4$-invariant VEV, they are explicitly written out as
\begin{eqnarray}
\chi_{11}^2+\chi_{11} \chi_{33}+ \chi_{33}^2+\chi_{12}^2+\chi_{13}^2+\chi_{23}^2 - \frac{\mu_\chi^2}{2g_1} &=& 0 \,; \label{eq:w_chi_singlet_v2} \\
\textstyle v_{\xi} \chi_{11} \chi_{23} \left( \frac{72 \sqrt{6}}{7} g_3-2 \sqrt{\frac{2}{3}} g_2\right)+v_{\xi} \chi_{12} \chi_{13} \left(2 \sqrt{\frac{2}{3}} g_2+\frac{96\sqrt{6}}{7} g_3\right) &=& 0 \,, \nonumber\\
\textstyle v_{\xi} (\chi_{11} + \chi_{33}) \chi_{13} \left(2 \sqrt{\frac{2}{3}} g_2-\frac{72 \sqrt{6}}{7} g_3\right)+v_{\xi} \chi_{12} \chi_{23} \left(2 \sqrt{\frac{2}{3}} g_2+\frac{96 \sqrt{6}}{7}  g_3\right) &=& 0 \,, \nonumber\\
\textstyle v_{\xi} \chi_{12} \chi_{33} \left(\frac{72 \sqrt{6}}{7}g_3-2 \sqrt{\frac{2}{3}} g_2\right)+v_{\xi} \chi_{13} \chi_{23} \left(2 \sqrt{\frac{2}{3}} g_2+\frac{96 \sqrt{6}}{7} g_3\right) &=& 0 \,; \label{eq:w_chi_triplet_v2}\\
\textstyle -g_4 \sqrt{\frac{2}{3}} v_{\xi} (\chi_{11}+2 \chi_{33}) &=& 0 \,, \nonumber\\
\textstyle g_4 \sqrt{\frac{2}{3}} v_{\xi} (2 \chi_{11}+\chi_{33}) &=& 0 \,. \label{eq:w_chi_quintet_v2}
\end{eqnarray}
Eq.~\eqref{eq:w_chi_quintet_v2} leads to $\chi_{11} = \chi_{33} = 0$. Taking it to Eq.~\eqref{eq:w_chi_triplet_v2}, we are left with $\chi_{12}\chi_{13} = \chi_{12}\chi_{23} = \chi_{13}\chi_{23}=0$, and therefore two of $\chi_{12},\chi_{13}, \chi_{23}$ vanishing. The only non-vanishing one is determined by Eq.~\eqref{eq:w_chi_singlet_v2}. All solutions are listed here, 
\begin{eqnarray}
\left(
\begin{array}{c}
 \langle \chi_{11} \rangle \\
 \langle \chi_{12} \rangle \\
 \langle \chi_{13} \rangle \\
 \langle \chi_{23} \rangle \\
 \langle \chi_{33} \rangle \\
\end{array}
\right)
= \left\{
\left(
\begin{array}{c}
 0 \\
 0 \\
 0 \\
 \pm \frac{v_\chi}{\sqrt{2}} \\
 0 \\
\end{array}
\right) \,, \ \ \ 
\left(
\begin{array}{c}
 0 \\
 0 \\
 \pm \frac{v_\chi}{\sqrt{2}} \\
 0 \\
 0 \\
\end{array}
\right) \,, \ \ \ 
\left(
\begin{array}{c}
 0 \\
 \pm \frac{v_\chi}{\sqrt{2}} \\
 0 \\
 0 \\
 0 \\
\end{array}
\right) 
\right\}\,.
\end{eqnarray}
Representing $\chi_{ij}$ by $\chi'$, $\chi''$ and $\chi_{1,2,3}$ in Eq.~\eqref{eq:rep_chi}, we obtain the result in Eq.~\eqref{eq:chi_VEV}.

\section{Deviation from the $Z_2$-invariant vacuum \label{app:Z2_correction}}
The naive estimation only gives the upper bound of the correction. The true correction may be smaller than it. It happens for the correction to the VEV of $\chi$. The minimisation of the superpotential including subleading higher dimensional operators is given by 
\begin{eqnarray}
&&\frac{g_2}{\Lambda}(\delta_\xi (\chi^{Z_2} \chi^{Z_2})_{\underline{5}})_{\underline{3}} + \frac{2 g_2}{\Lambda}(\xi^{A_4} (\delta_\chi \chi^{Z_2})_{\underline{5}})_{\underline{3}} + 
\frac{g_3}{\Lambda}(\delta_\xi (\chi^{Z_2} \chi^{Z_2})_{\underline{9}})_{\underline{3}} + \frac{2g_3}{\Lambda}(\xi^{A_4} (\delta_\chi \chi^{Z_2})_{\underline{9}})_{\underline{3}}  \nonumber\\ 
&& \hspace{3cm}
\approx \frac{1}{\Lambda^2} (\xi^{A_4} \chi^{Z_2})_{\underline{3}} \bar{\eta}^2 +
\frac{1}{\Lambda^2} \varphi (\chi^{Z_2}  \chi^{Z_2} )_{\underline{1}} \bar{\eta} + \cdots \,, \nonumber\\
&&(\delta_\xi \chi^{Z_2})_{\underline{5}} + (\xi^{A_4} \delta_\chi)_{\underline{5}} \approx 
\frac{1}{\Lambda} (\varphi^{Z_3}  \chi^{Z_2} )_{\underline{5}} \bar{\eta} +
\frac{1}{\Lambda} \chi^{Z_2} (\zeta^{\mathbf{1}'}  \zeta^{\mathbf{1}'} )_{\underline{1}} \cdots \,.
\label{eq:minimisation_xi}
\end{eqnarray}
Ignoring all the other subleading operators, we calculate its correction in detail instead of using the naive estimation. In this case, Eq.~\eqref{eq:minimisation_xi} is simplified to  
\begin{eqnarray}
&&\frac{2 g_2}{\Lambda}(\xi^{A_4} (\delta_\chi \chi^{Z_2})_{\underline{5}})_{\underline{3}} + 
\frac{2g_3}{\Lambda}(\xi^{A_4} (\delta_\chi \chi^{Z_2})_{\underline{9}})_{\underline{3}}  
\approx \frac{1}{\Lambda^2} (\xi^{A_4} \chi^{Z_2})_{\underline{3}} \bar{\eta}^2 \,, \nonumber\\
&&(\xi^{A_4} \delta_\chi)_{\underline{5}} \approx 0 \,.
\label{eq:minimisation_xi2}
\end{eqnarray}
Here, we has ignored the correction to the $\xi$ VEV since it is too small as discussed in the above. The above equation is explicitly written out as 
\begin{eqnarray}
&&\begin{pmatrix}\left(\frac{72}{7}  g_{3} - \frac{2}{3}  g_{2}\right) (\delta_{\chi'}+\delta_{\chi''}) \\
\sqrt{\frac{2}{3}} \left(g_{2}+\frac{144}{7} g_{3}\right) \delta_{\chi_3}
\\
\sqrt{\frac{2}{3}} \left(g_{2}+\frac{144}{7} g_{3}\right) \delta_{\chi_2}
\end{pmatrix} 
\frac{v_\xi v_\chi}{\Lambda }
\approx 
\begin{pmatrix}
 1 \\
 0 \\
 0 \\
\end{pmatrix} 
\frac{v_\xi v_\chi v_{\bar{\eta }}^2}{\sqrt{3} \Lambda ^2} \,, \nonumber\\
&&
\begin{pmatrix}
 \delta_{\chi''}-\delta_{\chi'} & 0 & 0 \\
 0 & \omega \delta_{\chi''} - \omega ^2 \delta_{\chi'} & 0 \\
 0 & 0 & \omega ^2 \delta_{\chi''} - \omega \delta_{\chi'}  \\
\end{pmatrix} i \sqrt{\frac{2}{3}} v_\xi 
\approx 0 \,.
\end{eqnarray}
This equation cannot give a self-consistent solution for $\delta_{\chi'}$ or $\delta_{\chi''}$ since the first equation predict $(\delta_{\chi'}+\delta_{\chi''})/v_\chi \sim v_{\bar{\eta}}^2/(\Lambda v_\chi)$ and the second one gives $\delta_{\chi'}/v_\chi \sim \delta_{\chi''}/v_\chi \sim 0$. It means that after subleading higher dimensional operators are included in the flavon superpotential,  $\partial w_f / \partial \chi^d_{\underline{3}} = 0$ and $\partial w_f / \partial \chi^d_{\underline{5}}= 0$ cannot hold at the same time. In other word, there is no flat direction for the flavon. 

Without flat direction, one has to calculate the VEV correction via the minimisation of the flavon potential. For similar discussion in only non-Abelian discrete symmetry, see e.g., Ref. \cite{deMedeirosVarzielas:2017hen}. 
In the model discussed here, the flavon potential is given by
\begin{eqnarray}
V_f &=& \left| \frac{\partial w_f}{\partial \chi^d_{\underline{3}}} \right|^2 + \left| \frac{\partial w_f}{\partial \chi^d_{\underline{5}}} \right|^2 + \cdots \,.
\end{eqnarray}
Taking the superpotential terms in Table~\ref{tab:flavon_superpotential} to $V_f$, we see that the first term is much smaller than the second term, $\left| \frac{\partial w_f}{\partial \chi^d_{\underline{3}}} \right|^2 \ll \left| \frac{\partial w_f}{\partial \chi^d_{\underline{5}}} \right|^2$. Therefore, the minimisation of $V_f$ is approximate to $\partial w_f / \partial \chi^d_{\underline{5}} = 0$, and the correction is given by 
\begin{eqnarray}
\frac{\delta_\chi}{v_\chi} 
&\sim & \max \{ \frac{\delta_{\xi}}{v_\xi}, \frac{v_\varphi v_{\bar{\eta}}}{\Lambda v_\xi}, \cdots \} 
= \frac{v_\varphi v_{\bar{\eta}}}{\Lambda v_\xi} \,. 
\end{eqnarray}

\end{document}